\documentclass[useAMS,usenatbib]{mnras}

\usepackage{ulem}
\usepackage{color}
\usepackage{graphicx}
\usepackage{times} 
\usepackage{amssymb}
\usepackage{amsmath}
\usepackage{natbib}
\usepackage{url}
\usepackage{multirow}
\usepackage{ctable}
\usepackage{threeparttable}
\newif\ifAMStwofonts
\AMStwofontstrue

\def\xmm{{\it XMM-Newton}}

\def\suzaku{{\it Suzaku}}

\def\chandra{{\it Chandra}}
\def\swift{{\it Swift}}

\def\epicpn{{EPIC-pn}}
\def\epicmos1{{EPIC-MOS1}}
\def\epicmos2{{EPIC-MOS2}}
\def\epicmos{{EPIC-MOS}}

\def\nustar{{\it NuSTAR}}
\def\hitomi{{\it Hitomi}}

\def\pcmsq{\hbox{$\rm\thinspace cm^{-2}$}}
\def\H0{{\rm ~km~s^{-1}~Mpc^{-1}}}

\def\kev{\hbox{\rm keV}}

\def\ergcms{{\rm ~erg~cm^{-2}~s^{-1}}}

\def\ergpcmsqps{\hbox{erg~cm$^{-2}$~s$^{-1}$}}

\def\ergps{\hbox{erg~s$^{-1}$}}

\def\msun{\hbox{$M_{\odot}$}}

\def\gps{g~s$^{-1}$}

\def\chisq{{$\chi^{2}$}}

\def\xspecv{\hbox{\small XSPEC}\, v12.6.0f}

\def\ciao{\hbox{\rm{\small CIAO}}}

\def\nustardas{\rm {\small NUSTARDAS}}

\def\addascaspec{\hbox{\rm{\small ADDASCASPEC~\/}}}
\def\flx2xsp{\rm{\small FLX2XSP}}

\def\epchain{\hbox{\rm{\small EPCHAIN}}}
\def\emchain{\hbox{\rm{\small EMCHAIN}}}

\def\rmfgen{\hbox{\rm{\small RMFGEN}}}
\def\arfgen{\hbox{\rm{\small ARFGEN}}}

\def\addascaspec{\hbox{\rm{\small ADDASCASPEC}}}
\def\nupipeline{\rm{\small NUPIPELINE}}
\def\nuproducts{\rm{\small NUPRODUCTS}}

\def\specextract{\hbox{\rm {\small SPECEXTRACT}}}

\def\grid25{\hbox{\rm{\small GRID25}}}

\def\simpl{{\small SIMPL}}

\def\tbabs{\rm{\small TBABS}}
\def\tbnew{\rm{\small TBNEW}}

\def\bb{{\small BB}}
\def\diskbb{{\small DISKBB}}
\def\diskpbb{{\small DISKPBB}}

\def\cutoffpl{{\small CUTOFFPL}}
\def\fdcut{{\small FDCUT}}
\def\npex{{\small NPEX}}

\def\comptt{\rm{\small COMPTT}}

\def\mekal{\rm{\small MEKAL}}

\def\ka{$\rm{K}\alpha$}

\def\eg{{\it e.g.}}

\def\ie{{\it i.e.~\/}}

\def\la{\mathrel{\hbox{\rlap{\hbox{\lower4pt\hbox{$\sim$}}}{\raise2pt\hbox{$<$}}}}}
\def\ga{\mathrel{\hbox{\rlap{\hbox{\lower4pt\hbox{$\sim$}}}{\raise2pt\hbox{$>$}}}}}

\def\d25{D$_{25}$}

\def\.25{0.25 keV\thinspace}

\def\rg{$R_{\rm G}$}
\def\rin{$R_{\rm in}$}

\def\rsp{$R_{\rm{sp}}$}
\def\rmag{$R_{\rm{M}}$}
\def\rco{$R_{\rm{co}}$}

\def\p13{NGC\,7793 P13}
\def\nsims{10,000}

\title[Broadband Spectroscopy of NGC\,7793 P13]{Super-Eddington Accretion onto
the Neutron Star NGC\,7793 P13: Broadband X-ray Spectroscopy and
Ultraluminous X-ray Sources}

\author[D.\,J. Walton et al.]
{\parbox{7.in}{D.\,J. Walton$^{1}$\thanks{E-mail: dwalton@ast.cam.ac.uk},
F. F\"urst$^{2}$,
F. A. Harrison$^{3}$,
D. Stern$^{4}$,
M. Bachetti$^{5}$,
D. Barret$^{6, 7}$, \\
M. Brightman$^{3}$,
A. C. Fabian$^{1}$,
M. J. Middleton$^{8}$,
A. Ptak$^{9}$,
L. Tao$^{3}$ \\
\\[-0.2cm]
\footnotesize
$^{1}$ \it{Institute of Astronomy, Cambridge University, Madingley Road, Cambridge, CB3 0HA} \\ 
$^{2}$ \it{European Space Astronomy Centre (ESA/ESAC), Operations Department, Villanueva de la Ca\~nada (Madrid), Spain} \\
$^{3}$ \it{Space Radiation Laboratory, California Institute of Technology, Pasadena, CA 91125, USA} \\
$^{4}$ \it{Jet Propulsion Laboratory, California Institute of Technology, Pasadena, CA 91109, USA} \\
$^{5}$ \it{INAF/Osservatorio Astronomico di Cagliari, via della Scienza 5, I-09047 Selargius (CA), Italy} \\
$^{6}$ Universite de Toulouse; UPS-OMP; IRAP; Toulouse, France \\
$^{7}$ CNRS; IRAP; 9 Av. colonel Roche, BP 44346, F-31028 Toulouse cedex 4, France \\
$^{8}$ \it{Department of Physics and Astronomy, University of Southampton, Highfield, Southampton SO17 1BJ, UK} \\
$^{9}$ \it{NASA Goddard Space Flight Center, Greenbelt, MD 20771, USA}}}

\date{}

\begin{document}
\pagerange{\pageref{firstpage}--\pageref{lastpage}}
\maketitle
\label{firstpage}

\begin{abstract}
We present a detailed, broadband X-ray spectral analysis of the ULX pulsar \p13, a
known super-Eddington source, utilizing data from the \xmm, \nustar\ and \chandra\
observatories. The broadband \xmm+\nustar\ spectrum of P13 is qualitatively similar
to the rest of the ULX sample with broadband coverage, suggesting that additional
ULXs in the known population may host neutron star accretors. Through
time-averaged, phase-resolved and multi-epoch studies, we find that two non-pulsed
thermal blackbody components with temperatures $\sim$0.5 and 1.5\,keV are required
to fit the data below 10\,keV, in addition to a third continuum component which extends
to higher energies and is associated with the pulsed emission from the accretion
column. The characteristic radii of the thermal components appear to be comparable,
and are too large to be associated with the neutron star itself, so the need for two
components likely indicates the accretion flow outside the magnetosphere is complex.
We suggest a scenario in which the thick inner disc expected for super-Eddington
accretion begins to form, but is terminated by the neutron star's magnetic field soon
after its onset, implying a limit of $B \lesssim 6 \times 10^{12}$\,G for the dipolar
component of the central neutron star's magnetic field. Evidence of similar termination
of the disc in other sources may offer a further means of identifying additional neutron
star ULXs. Finally, we examine the spectrum exhibited by P13 during one of its unusual
`off' states. These data require both a hard powerlaw component, suggesting residual
accretion onto the neutron star, and emission from a thermal plasma, which we argue is
likely associated with the P13 system.
\end{abstract}


\section{Introduction}
\label{sec_intro}

The ultraluminous X-ray source (ULX\footnote{Here we define a ULX to be an
off-nuclear X-ray point source with a luminosity in excess of $10^{39}$\,\ergps\ in the
0.3--10.0\,keV bandpass, assuming isotropy.}) \p13\ (hereafter P13), which reaches
luminosities of $L_{\rm{X,peak}} \sim 10^{40}$\,\ergps, was recently found to be
powered by an accreting neutron star through the detection of coherent X-ray
pulsations (\citealt{Fuerst16p13, Israel17p13}). Along with M82\,X-2, the first ULX
pulsar discovered ($L_{\rm{X, peak}} \sim 2 \times 10^{40}$\,\ergps;
\citealt{Bachetti14nat}) and NGC\,5907 ULX ($L_{\rm{X, peak}} \sim 9 \times
10^{40}$\,\ergps; \citealt{Israel17, Fuerst17ngc5907}), only three neutron star ULXs
with such extreme luminosities are currently known. These systems are of great interest,
as they offer a rare opportunity to study accretion in the highly super-Eddington regime
(the Eddington limit for a standard 1.4\,\msun\ neutron star is $L_{\rm{E}} \sim 2 \times
10^{38}$ \ergps).


There are several observational similarities seen between the P13, M82 X-2 and
NGC\,5907 ULX systems. All three have broadly similar pulse periods, of order
$\sim$1\,s (\citealt{Bachetti14nat, Fuerst16p13, Israel17, Israel17p13}), and exhibit
strong secular spin-up owing to the large accretion torques related to their extreme
accretion rates; over just $\sim$10 years, the pulse period in NGC\,5907 ULX
evolved from $\sim$1.4 to $\sim$1.1\,s. The pulse profiles observed are also similar,
with all three sources showing broad, near-sinusoidal variations. Furthermore, all
three show long-term periodicities ($\sim$60--80\,d; \citealt{Motch14nat,
Walton16period, Kong16, Hu17}, Brightman et al. \textit{in prep.}). In the cases of
M82 X-2 and NGC\,5907 ULX these periods are known to be super-orbital, as
variations in their pulse periods have also revealed orbital periods of a few days. No
such independent constraints on the orbit are currently available for P13, where the
nature of the $\sim$64\,d period remains uncertain, although it does seem to show
some `jitter' (\citealt{Motch14nat}) that may be indicative of super-orbital variations.
Finally, all three sources show unusual `off-states', where the X-ray flux is diminished
by a factor of $\sim$50 or more (\eg\ \citealt{Motch14nat, Walton15,
Brightman16m82a}). \cite{Tsygankov16} suggest that these extreme flux modulations
may be related to the propeller effect, but their nature is not currently well established,
and may even differ between systems.

These systems pose key questions, particularly with regards to how these neutron
stars are able to reach such extreme apparent luminosities. The magnetic field of the
central neutron star will channel the accretion flow into columns inside the
magnetospheric radius (\rmag, the point at which magnetic pressure dominates),
allowing material to accrete onto the magnetic poles while radiation escapes from the
sides of the column \citep{Basko76}. This introduces anisotropy into the radiation field
(which is required in order for X-ray pulsations to be observed), and in turn a beaming
correction to simple luminosity estimates becomes necessary. However, while some
anisotropy is clearly required, it is difficult to explain the near-sinusoidal pulse profiles
observed from all three systems in the context of strong beaming of sub-Eddington
accretion onto an otherwise standard neutron star (which would require beaming by
factors of  $\sim$100 or more).

Some authors have invoked strong, magnetar-level magnetic fields ($B \sim
10^{13-14}$\,G; \eg\ \citealt{Eksi15, DallOsso15, Mushtukov15}). This reduces the
scattering cross section for electrons \citep{Herold79}, reducing the radiation pressure
and in turn increasing the effective Eddington luminosity. However, other authors have
instead suggested that the magnetic field is much lower (potentially as low as $B \sim
10^9$\,G) based on the ratio of the spin-up rate to the luminosity, which is an order of
magnitude lower than typical X-ray pulsars (\eg\ \citealt{Kluzniak15}). These authors
argue that a disk truncated at a large radius (close to the co-rotation radius, \rco, as
would occur for a high-B field system; \rco\ is the point at which the material in the disc
co-rotates with the neutron star, and \rmag\ $<$ \rco\ is required for strong accretion to
occur) would not provide the required lever arm to power the observed spin up. In such
a scenario, the extreme luminosities would need to be produced by a highly
super-Eddington accretion disk that  extends close to the accretor, similar to
super-Eddington accretion onto a black hole (\citealt{King16ulx}).

P13 provides a key laboratory for understanding these extreme neutron stars and
their link to the broader ULX population. With only three examples, the sample of
ULX pulsars is still extremely limited, and there are obstacles to undertaking detailed analyses with M82 X-2 and NGC\,5907 ULX. M82 X-2 is heavily confused with its
more luminous neighbour M82 X-1 (separated by $\sim$5$''$), and NGC\,5907 ULX
is situated at a large distance ($\sim$17\,Mpc; \citealt{Tully16}) so its observed flux is
relatively low. Both sources are also heavily absorbed ($N_{\rm{H}} \sim 10^{22}$
cm$^{-2}$), severely complicating comparisons with other ULXs. In contrast, P13 is
both well isolated and relatively unobscured ($N_{\rm{H}} \sim 10^{21}$\,cm$^{-2}$).
Furthermore, pulsations are only intermittently detected in M82 X-2 and NGC\,5907
ULX (\citealt{Bachetti14nat, Israel17}), while for P13 they have always been
detected when the source has been observed at high flux and with sufficient temporal
resolution.

In this paper, we present results from the 2016 broadband observation of P13 with
\xmm\ (\citealt{XMM}) and \nustar\ (\citealt{NUSTAR}). Our initial work on these data,
presented in \cite{Fuerst16p13}, focused on the detection and evolution of the X-ray
pulsations, and here we focus on undertaking a comprehensive spectral analysis.
We describe the observations and data reduction in Section \ref{sec_red}, and we
provide details on our spectral analysis in Sections \ref{sec_2016} and \ref{sec_ME}.
We draw comparisons with the broader ULX population in Section \ref{sec_ulxpop},
and in Section \ref{sec_dis} we discuss the results. Finally, we summarise our
conclusions in Section \ref{sec_conc}. Throughout this work, we adopt a distance to
NGC\,7793 of 3.5\,Mpc (\citealt{Pietrzynski10}).

\begin{table}
  \caption{Details of the X-ray observations of NGC\,7793 P13 considered in this work}
\begin{center}
\vspace{-0.4cm}
\begin{tabular}{c c c c c}
\hline
\hline
\\[-0.2cm]
Epoch & Mission(s) & OBSID(s) & Start & Exposure\tmark[a] \\
\\[-0.3cm]
& & & Date & (ks) \\
\\[-0.3cm]
\hline
\hline
\\[-0.2cm]
\multicolumn{5}{c}{\textit{2016}} \\
\\[-0.3cm]
\multirow{2}{*}{XN1} & \nustar\ & 80201010002 & \multirow{2}{*}{2016-05-20} & 115\\
& \xmm\ & 0781800101 & & 28/46 \\
\\[-0.2cm]
\multicolumn{5}{c}{\textit{Archival Data}} \\
\\[-0.3cm]
C1 & \chandra\ & 3954 & 2003-09-06 & 49 \\
\\[-0.3cm]
C2 & \chandra\ & 14231 & 2011-08-13 & 59 \\
\\[-0.3cm]
C3 & \chandra\ & 13439 & 2011-12-25 & 58 \\
\\[-0.3cm]
C4 & \chandra\ & 14378 & 2011-12-30 & 25 \\
\\[-0.3cm]
X1 & \xmm\ & 0693760101 & 2012-05-14 & 26/36 \\
\\[-0.3cm]
X2 & \xmm\ & 0693760401 & 2013-11-25 & 41/47 \\
\\[-0.3cm]
X3 & \xmm\ & 0748390901 & 2014-12-10 & 42/48 \\
\\[-0.3cm]
\hline
\hline
\\[-0.4cm]
\end{tabular}
\end{center}
$^{a}$ \xmm\ exposures are listed for the \epicpn/MOS detectors 
\label{tab_obs}
\end{table}

\begin{figure*}
\begin{center}
\hspace*{-0.5cm}
\rotatebox{0}{
{\includegraphics[width=235pt]{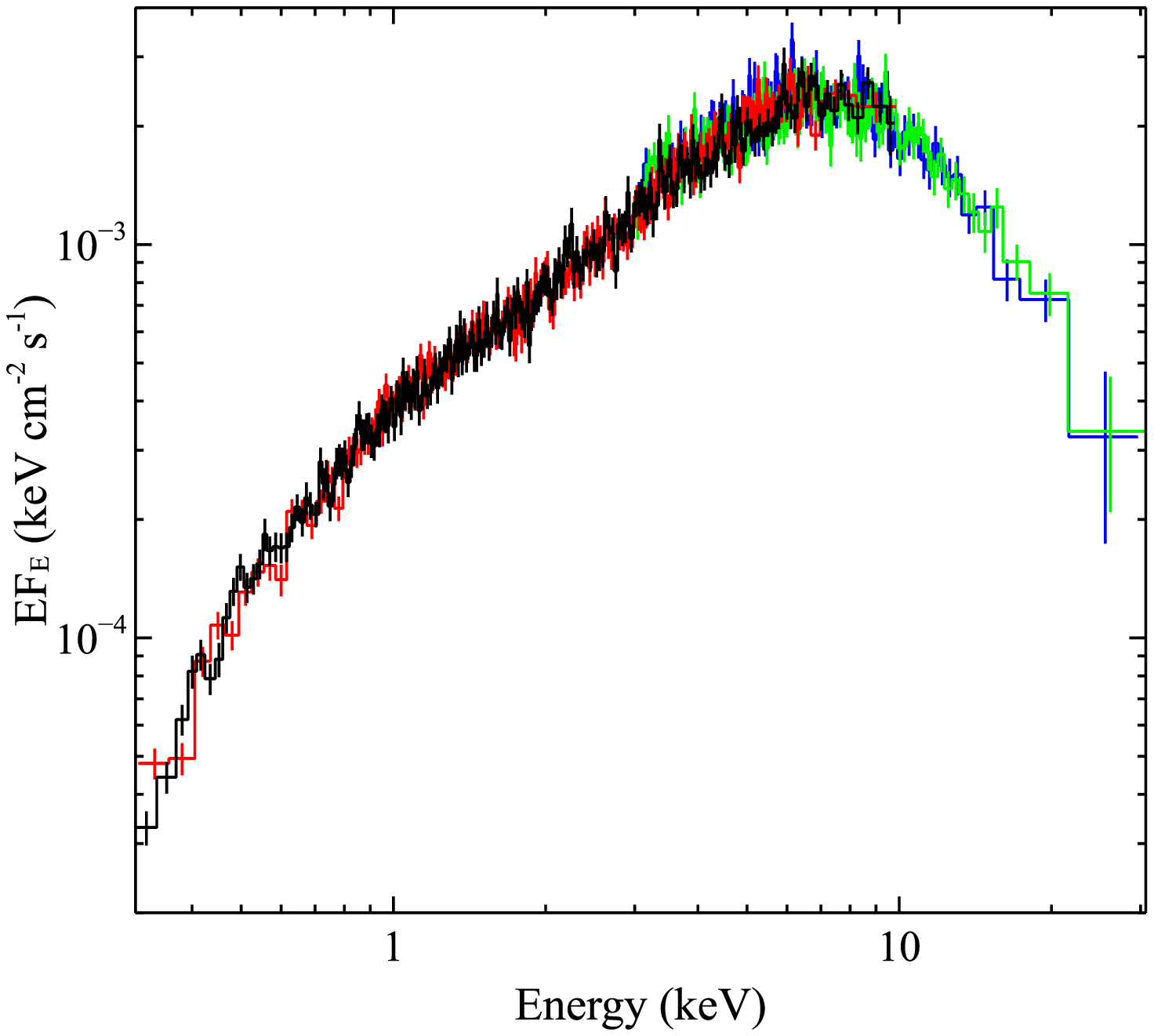}}
}
\hspace*{0.5cm}
\rotatebox{0}{
{\includegraphics[width=235pt]{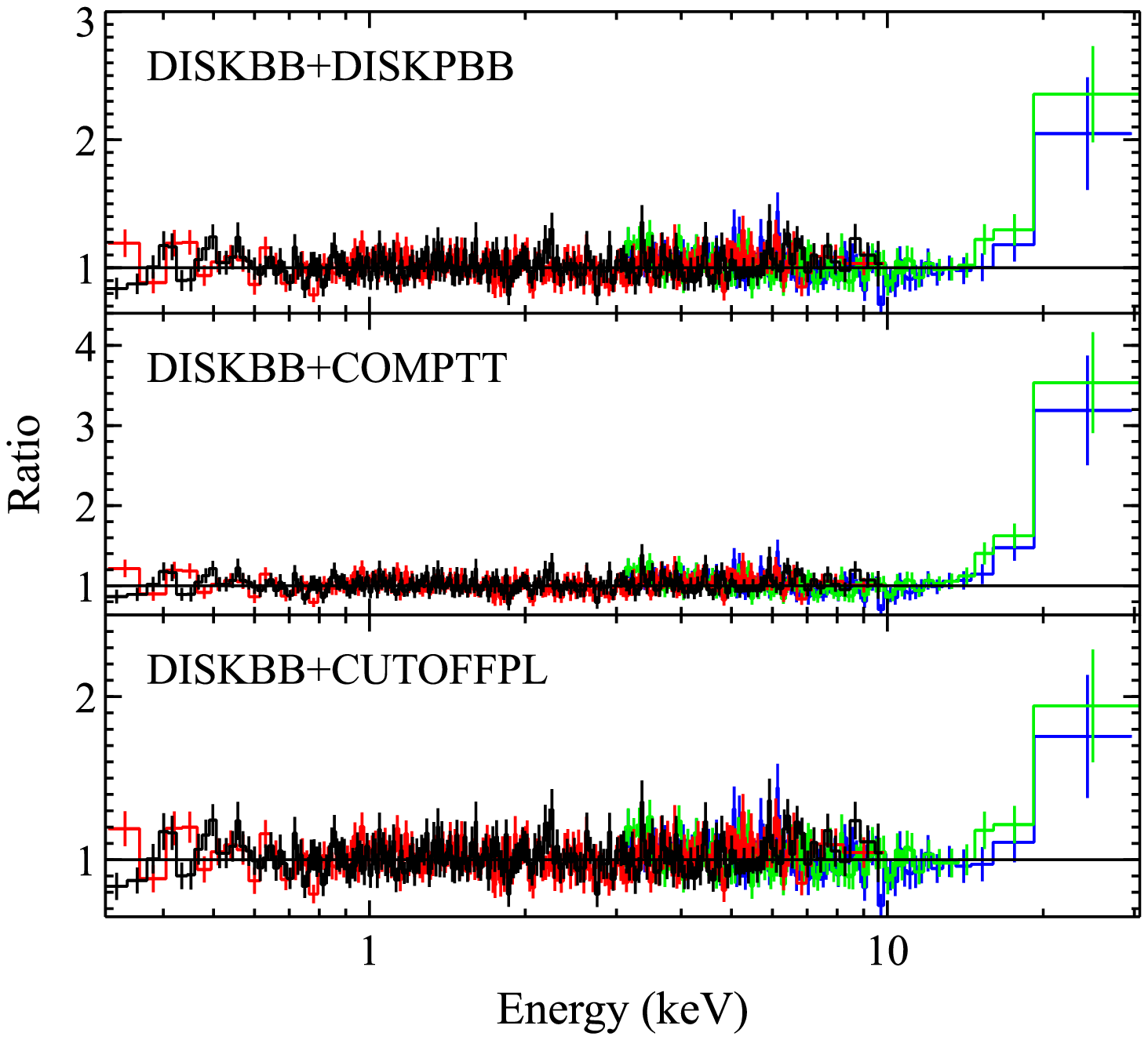}}
}
\end{center}
\caption{
\textit{Left panel:} the time-averaged broadband X-ray spectrum of \p13, observed
by \xmm\ (\epicpn\ in black, \epicmos\ in red) and \nustar\ (FPMA in green, FPMB in
blue), unfolded through a model simply consisting of a constant (formally, the
model used is a powerlaw with $\Gamma = 0$). The \nustar\ data clearly demonstrate
the presence of a high-energy spectral cutoff. \textit{Right panel:} data/model ratios for
the three continuum models initially considered (see Section \ref{sec_avspec}). Each
results in a clear excess in the residuals at high energies, indicating the presence of
an additional high-energy continuum component. The data have been rebinned for
visual clarity.}
\label{fig_spec}
\end{figure*}


\section{Observations and Data Reduction}
\label{sec_red}

As discussed in \cite{Fuerst16p13}, the coordinated \xmm+\nustar\ observation of
P13 that is the focus of much of this work was triggered following the detection of a
high flux with \swift\ (\citealt{SWIFT}), and currently represents the highest flux
probed with high signal-to-noise (S/N) observations for this source to date. In addition
to these data, we also consider the archival \xmm\ and \chandra\ (\citealt{CHANDRA})
exposures of P13. Details of all the observations considered in this work are given in
Table \ref{tab_obs}. Our data reduction procedure largely follows that outlined in
\cite{Fuerst16p13}, but here we repeat the salient points for clarity.

\subsection{\textit{NuSTAR}}

The \nustar\ data were reduced using the \nustar\ Data Analysis Software ({\small
NUSTARDAS}; v1.6.0) following standard procedures; \nustar\ caldb v20160824 was
used throughout. We cleaned the unfiltered event files with \nupipeline, applying the
standard depth correction (which significantly reduces the internal background) and
removing passages through the South Atlantic Anomaly. Barycentre corrections were
applied to the photon arrival times in the cleaned event files using using the DE200
reference frame, and based on the pulse period reported in \cite{Fuerst16p13} we
phase-stamp each of the events. Source products and instrumental response files
were extracted from circular regions of radius 70$''$ for both of the focal plane
modules (FPMA/B) with \nuproducts, with background estimated from a much larger
region on the same detector as P13. In addition to the standard `science' data, to
maximize the signal-to-noise (S/N) we also extract the `spacecraft science' data
following the procedure outlined in \cite{Walton16cyg}, which provides $\sim$10\% of
the good exposure in this case.

\subsection{\textit{XMM-Newton}}

The \xmm\ data were reduced with the \xmm\ Science Analysis System (v15.0.0),
following the standard data reduction guide.\footnote{http://xmm.esac.esa.int/}.
The raw data files were cleaned using \epchain\ and \emchain\ for the \epicpn\
(\citealt{XMM_PN}) and each of the \epicmos\ (\citealt{XMM_MOS}) detectors,
respectively. As with the \nustar\ data, barycentre corrections were applied to the
photon arrival times in the cleaned event files for \epicpn\ using the DE200 reference
frame, and we phase-stamp events based on the pulse periods reported in
\cite{Fuerst16p13}; unfortunately the \epicmos\ detectors do not have sufficient
temporal capabilities to resolve the $\sim$0.4\,s pulse period. Source products were
typically extracted from circular regions of radius $\sim$40$''$ and background was
estimated from larger areas on the same CCD free of other contaminating point
sources. The exception was OBSID 0693760101, during which the source was
extremely faint (but still clearly detected), so a radius of 20$''$ was used. As
recommended, we only use single--double patterned events for \epicpn\ and
single--quadruple patterned events for \epicmos, and we also excluded periods of
high background as standard. Instrumental redistribution matrices and ancillary
response files were generated for each of the detectors with \rmfgen\ and \arfgen,
respectively. Finally, after checking their consistency, spectra from the two \epicmos\
detectors were combined using \addascaspec.

\begin{table*}
  \caption{Best fit parameters obtained for the broadband continuum models
  applied to the P13 data from epoch XN1}
\begin{center}
\begin{tabular}{c c c c c c c c c c}
\hline
\hline
\\[-0.2cm]
Model\tmark[a] & $N_{\rm H; int}$ & $kT_{\rm{in; DBB}}$ & Norm & $p / \tau / \Gamma_{\rm{CPL}}$ & $kT\tmark[b] / E_{\rm{cut}}$ & Norm & $\Gamma_{\rm{SIMPL}}$ & $f_{\rm scat}$ & $\chi^{2}$/DoF \\
\\[-0.3cm]
& [$10^{20}$ cm$^{-2}$] & [\kev] & & & [\kev] & & & [\%] & \\
\\[-0.3cm]
\hline
\hline
\\[-0.2cm]
\diskpbb\ & $8 \pm 1$ & $0.45^{+0.03}_{-0.04}$ & $1.1^{+0.4}_{-0.2}$ & $>1.3$\tmark[c] & $2.0 \pm 0.3$ & $8^{+6}_{-5} \times 10^{-2}$ & $3.7^{+0.9}_{-1.1}$ & $>19$ & 1129/1148 \\
\\[-0.2cm]
\comptt\ & $8 \pm 1$ & $0.49^{+0.10}_{-0.20}$ & $1.0^{+2.3}_{-0.5}$ & $17 \pm 2$ &  $1.4^{+0.3}_{-0.1}$ & $7.0^{+0.6}_{-1.0} \times 10^{-4}$ & $4.0^{+0.3}_{-0.7}$ & $>36$ & 1129/1148 \\
\\[-0.2cm]
\cutoffpl\ & $8 \pm 1$ & $0.48^{+0.08}_{-0.05}$ & $1.0 \pm 0.4$ & $-0.9^{+0.4}_{-0.6}$ & $1.9^{+0.7}_{-0.6}$ & $(2.5 \pm 0.5) \times 10^{-4}$ & $4.0^{+0.7}_{-1.8}$ &  $>12$ & 1129/1148 \\
\\[-0.3cm]
\hline
\hline
\\[-0.4cm]
\end{tabular}
\label{tab_param_av}
\end{center}
\flushleft
$^a$ The base continuum models fit here are e.g. TBNEW$_{\rm{Gal}} \times$ TBNEW$_{\rm{int}} \times ($DISKBB $+$ CUTOFFPL $\otimes$ SIMPL$)$ \\
$^b$ Here, $kT$ refers to either the inner temperature for the DISKPBB model, or the electron temperature for the COMPTT model \\
$^c$ During the fitting process, we restrict the radial temperature index to $0.5 \leq p \leq 2.0$
\end{table*}

\subsection{\textit{Chandra}}

The \chandra\ data were reduced with the \ciao\ software package (v4.7) following
standard procedure. We extracted spectra from the ACIS-S detector
(\citealt{CHANDRA_ACIS}) level 2 event files and computed instrumental response
files using \specextract. Although the three 2011 observations all occurred during
the extended off-state seen from P13, the source is still clearly detected in all
observations. Source spectra are extracted from circular regions of radius 4--5$''$,
and as with both \nustar\ and \xmm\ background was estimated from larger areas
on the same CCD free of other contaminating point sources. All the observations
were performed in the Timed Exposure mode with fixed frame times of $\sim$3.2\,s;
the time-resolution of these \chandra\ observations is therefore too coarse to detect
the $\sim$0.4\,s pulsations from P13.

\section{NGC\,7793 P13: The Broadband 2016 Dataset}
\label{sec_2016}

We begin our analysis by focusing on the broadband \xmm+\nustar\ dataset obtained
in 2016 (\ie epoch XN1), as these observations have provided the first ever hard X-ray
($>$10\,keV) detection of P13. Model fits are performed with \xspecv\ (\citealt{xspec}),
parameter uncertainties are quoted at the 90\% confidence level for one interesting
parameter ($\Delta\chi^{2} = 2.7$, with all other free parameters allowed to vary
during the error estimation). All models considered include a Galactic absorption
component with a fixed column of $N_{\rm{H,Gal}} = 1.2 \times 10^{20}$\,\pcmsq\
(\citealt{NH}), and we also allow for absorption intrinsic to the source at the redshift of
NGC\,7793 ($N_{\rm{H; int}}$; $z = 0.000767$). Both neutral absorption components
are modelled with the \tbnew\ absorption code. As recommended for \tbnew, we use the
cross-sections of \cite{Verner96} and the abundance set presented by \cite{tbabs}. We
also allow for cross-calibration uncertainties between the different detectors by including
multiplicative constants that are allowed to float between the datasets, fixing \epicpn\
at unity. These constants are always within 10\% of unity, as expected
(\citealt{NUSTARcal}). Unless stated otherwise, spectra are grouped to have a
minimum of 50 counts per energy bin throughout this work in order to facilitate the use
of $\chi^{2}$ minimization during our analysis.

\subsection{Time-Averaged Spectroscopy}
\label{sec_avspec}

The time-averaged broadband spectrum observed from epoch XN1 is shown in Figure
\ref{fig_spec} (\textit{left panel}); P13 is detected by \nustar\ up to $\sim$30\,keV. The
broadband spectrum is qualitatively similar to the broadband spectra seen from the rest
of the ULX population to date (see Section \ref{sec_ulxpop}). As noted by
\cite{Motch14nat} and \cite{Pintore17}, who present spectral analyses of the archival
\xmm\ observations of P13, the spectrum below 10\,keV requires two broadband
continuum components: a thermal component that contributes below $\sim$2\,keV, and
a harder component that dominates at higher energies. The prior \xmm\ observations
analyzed by these authors indicate this higher energy component shows curvature in
the $\sim$5--10\,keV band, similar to the rest of the ULX population (\eg\
\citealt{Stobbart06, Gladstone09}). The \nustar\ data clearly and robustly confirms the
presence of this curvature, demonstrating that, as with the rest of the ULX sample
observed by \nustar\ to date, the emission that dominates the 2-10\,keV band is not a
high-energy powerlaw tail, and that the curvature seen in the \xmm\ data does mark
the start of a high-energy spectral cutoff.

Before undertaking a more detailed analysis of the spectral properties of P13,
we fit the average spectrum with a set of models typically applied to other ULXs in the
literature to provide a simple comparison. Based on the previous analyses of the
archival P13 data, and our prior work on broadband observations of ULXs with \nustar,
we apply a set of 2-component continuum models commonly applied to ULX spectra.
We start with a model consisting of two accretion disk components, combining the
\diskbb\ and \diskpbb\ models (\citealt{diskbb, diskpbb}). The former is based on the
thin accretion disc model of \cite{Shakura73}, which has a radial temperature index of
$p = 3/4$ (where $T(r) \propto r^{-p}$) and just has the disc temperature ($T_{\rm{in}}$)
as a free parameter, while the latter also allows $p$ to be varied as a free parameter.
Second, we also apply a model combining \diskbb\ with \comptt\ (\citealt{comptt}), a
thermal Comptonization model which is primarily characterized by the optical depth
($\tau$) and the temperature ($kT_{\rm{e}}$) of the scattering electrons (for
convenience, we assume the temperature of the seed photons to be that of the \diskbb\
component, as in previous works). Both of these models provide reasonable fits to the
data below $\sim$10--15\,keV. However, as can clearly be seen in the data/model ratios
shown in Figure \ref{fig_spec} (\textit{right panels}), both models leave a clear excess in
the \nustar\ data at the highest energies probed.

To further test the presence of this hard excess, we also fit a third model, combining
\diskbb\ with a more phenomenological \cutoffpl\ component. This is a simple
powerlaw continuum with an exponential cutoff ($E_{\rm{cut}}$), which has a broader
curvature at high energies than both of the \diskpbb\ and \comptt\ models discussed
above. This model has often been applied to ULX data (\eg \citealt{Bachetti13,
Pintore17}), but is also fairly representative of models typically applied to Galactic
neutron stars (\eg\ \citealt{Coburn02, Fuerst14})\footnote{We have also tried replacing
the \cutoffpl\ model with the \fdcut\ and \npex\ models also frequently applied to Galactic
neutron stars (\eg\ \citealt{Odaka13, Fuerst13, Islam15, Vybornov17}), but these provide
identical fits to the \cutoffpl\ model, so we do not present the results in full.}. However,
although it is weaker than both the previous models, even the broader curvature cannot
resolve the hard excess (see Figure \ref{fig_spec}); a third continuum component is
required to fit the time-averaged data regardless of the model used to fit the curvature in
the $\sim$5--10\,keV band. Similar hard excesses have now been seen in a number of
other ULXs observed with \nustar\ (\eg\ \citealt{Walton15hoII, Walton17hoIX,
Mukherjee15, Fuerst17ngc5907}). Following these works, to account for this excess
emission we add \simpl\ (\citealt{simpl}) to the higher energy component in the models
described above. This is a convolution model that scatters some fraction ($f_{\rm{scat}}$)
of the photons in an input continuum model into a high-energy powerlaw tail (\ie photon
number is conserved). The additional of \simpl\ resolves the hard excess and results in a
significant improvement to the fit for all three of the base continuum models considered;
even for the \diskbb+\cutoffpl\ model -- for which the hard excess is weakest -- the fit
improves by $\Delta\chi^{2} = 32$ for two additional free parameters.

In order to confirm that this additional high-energy component is required by the
data at a significant level we performed a series of simulations, using the
\diskbb+\cutoffpl\ continuum model to be conservative, as this is the model that returns
the lowest $\Delta\chi^{2}$ when this component is included in the fit. Using the same
response and background files, and adopting the same exposure times as the real data,
we simulated \nsims\ sets of \xmm\ (pn and combined MOS1+MOS2) and \nustar\
(FPMA and FPMB) spectra with the {\small FAKEIT} command in XSPEC based on the
best-fit \diskbb+\cutoffpl\ model. Each of the simulated datasets was rebinned in the
same manner and analysed over the same bandpass as the real data. We then fit each
of the simulated \xmm+\nustar\ datasets with the \diskbb+\cutoffpl, and noted the
$\Delta\chi^{2}$ improvement the addition of a \simpl\ component provided over this fit.
Of the \nsims\ datasets simulated, none returned a chance improvement equivalent to
or greater than that observed, implying that this component is seen in the real data with
a significance comfortably in excess of 3$\sigma$ (for reference, the false-alarm
probability for a 3$\sigma$ detection would correspond to $\Delta\chi^{2} \sim 9$ here).

The results for the three models considered here, which all provide statistically equivalent
fits, are presented in Table \ref{tab_param_av}. While the broadband spectrum is
qualitatively similar to the rest of the sample of ULXs observed by \nustar\ to date, there
are some notable quantitative differences. For example, the temperature of the lower
energy \diskbb\ component is $kT_{\rm{in; DBB}} \sim 0.5$\,keV, slightly hotter than for
other ULXs where the same models usually find $kT_{\rm{in; DBB}} \sim 0.3$\,keV.
Furthermore, the rise of the spectrum up to the peak of the emission at $\sim$7\,keV is
harder than typically observed from other ULXs. This can be seen in the results for the
radial temperature index in the \diskpbb\ model and the optical depth in the \comptt\
model; in the former case we find $p > 1.3$ while ULXs more typically show $p < 0.75$
when fit with \diskpbb, and in the latter case we find $\tau \sim 17$ while typical ULX
values are $\tau \sim 5-10$ with this model. This means the continuum component
dominating the 2-10\,keV emission is much more peaked (\ie the spectral curvature in
this band is stronger) in P13 than in the broader ULX population. Indeed, if we replace
this component with a single-temperature blackbody, a similarly good fit is obtained
($\chi^{2}$/DoF = 1133/1149), so a range of temperatures is not strongly required in
this case.

\subsection{The Pulsed Emission}
\label{sec_pulse}

To further characterize the spectral behaviour exhibited by P13, we also investigate
spectral variations across the pulse period in an attempt to separate the emission
from the accretion column (pulsed) and rest of the accretion flow (which for simplicity
is assumed to be steady over the pulse cycle). Owing to its superior time resolution,
in this section we are only able to utilize data from the \xmm\ \epicpn\ detector for the
lower energy data; the pulse period is too short to be resolved by the timing
capabilities of the \epicmos\ detectors.

\begin{figure}
\begin{center}
\hspace*{-0.6cm}
\rotatebox{0}{
{\includegraphics[width=235pt]{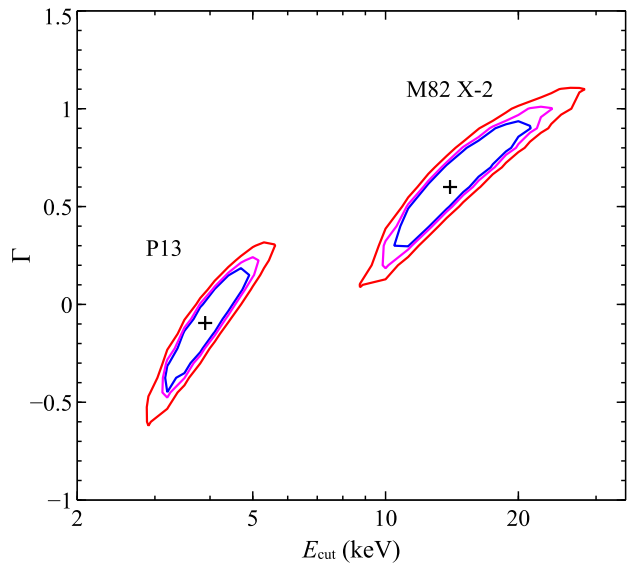}}
}
\end{center}
\caption{
2D confidence contours for $\Gamma$ and $E_{\rm{cut}}$ for the pulsed
spectra of P13 and M82\,X-2. The 90, 95 and 99\% confidence contours for 2
parameters of interest are shown in blue, magenta and red, respectively.}
\label{fig_pulse_gamEc}
\end{figure}


\subsubsection{Difference Spectroscopy}
\label{sec_pulsediff}

In order to undertake a comparison between P13 and M82 X-2, we compute
the spectrum of the pulsed component for the broadband \xmm+\nustar\ dataset. This
approach is necessary since the absolute spectrum of M82 X-2 cannot easily be
disentangled from the contribution from M82 X-1. We follow the approach taken for
M82 X-2 in \cite{Brightman16m82a}, and subtract the average spectrum obtained
during the faintest quarter (\ie $\Delta\phi_{\rm{pulse}} = 0.25$) of the pulse cycle from
that extracted during the brightest quarter (\ie ``pulse on"$-$``pulse-off") in order to
isolate the pulsed emission in a simple manner. The data are rebinned to have a
minimum S/N per energy bin of 3, to allow the use of \chisq\ statistics, and we fit the
data over the 0.3--20.0\,keV energy range with a simple \cutoffpl\ model. The
intrinsic column density is poorly constrained in these fits, and so we fix it to $8 \times
10^{20}$\,\pcmsq\ following our analysis of the average spectrum, and we also fix the
cross-calibration constants between \epicpn\ and FPMA/B to the best-fit values
obtained in that analysis given the low S/N of the pulsed spectrum. This provides a
good fit to the data, with $\chi^{2} = 113$ for 91 DoF. However, for P13 we find the
pulsed emission to have a much harder rise, and a lower cutoff energy: $\Gamma =
-0.1 \pm 0.3$, $E_{\rm{cut}} = 3.9^{+0.8}_{-0.6}$\,keV. Identical results are obtained if
we minimise the Cash statistic (\citealt{cstat}) instead. For comparison, we show the
2-D confidence contours for these parameters in Figure  \ref{fig_pulse_gamEc} for
both P13 and M82 X-2. The pulsed flux from P13 during this epoch corresponds to
an apparent 0.3--20.0\,keV luminosity of $(4.0\pm0.3) \times 10^{39}$\,\ergps\ were
one to assume isotropic emission (although clearly this is not the case).


\begin{figure}
\begin{center}
\hspace*{-0.6cm}
\rotatebox{0}{
{\includegraphics[width=235pt]{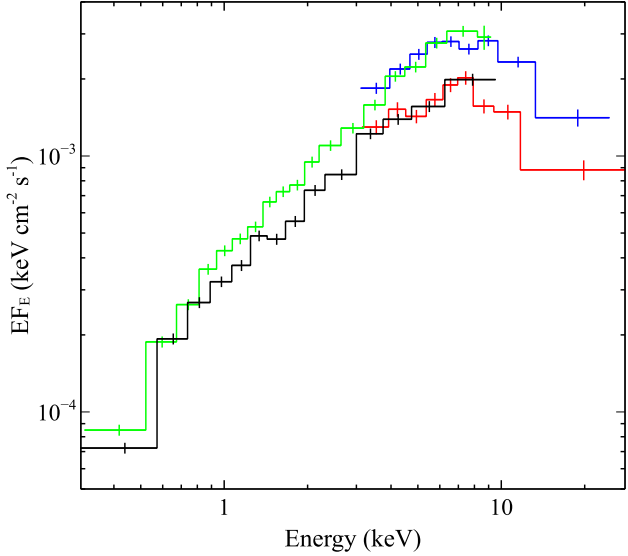}}
}
\end{center}
\caption{
The broadband X-ray spectrum of \p13 extracted from the peak and the
trough of the pulse cycle during epoch XN1. As with Figure \ref{fig_spec} (\textit{left
panel}), the data have been unfolded through a model consisting of a constant.
For clarity, the data have been strongly rebinned such that the differences are visually
apparent, and we only show the \xmm\ \epicpn\ (low: black, high: green) and \nustar\
FPMA (low: red, high: blue) data.}
\label{fig_spec_pulse}
\end{figure}


\subsubsection{Phase-Resolved Spectroscopy}
\label{sec_pulsephase}

In addition to the ``pulse-on"$-$``pulse-off" spectroscopy, we also fit the spectra
extracted across a series of phase bins. We continue using phase bins of
$\Delta\phi_{\rm{pulse}} = 0.25$ in size, and extract spectra from four periods
across the pulse cycle: peak, fall, trough and rise. However, on inspection the fall
and rise spectra were found to be practically identical, so we combined them to
form a higher S/N medium-flux spectrum, resulting in three broadband spectra to
be fit in our phase-resolved analysis: high-flux (peak), medium-flux (rise $+$ fall)
and low-flux (trough). We show the high- and low-flux spectra in Figure
\ref{fig_spec_pulse} for comparison. It is clear that the majority of the difference
between the two is seen at higher energies, which is not surprising since the pulse
fraction is known to increase with energy (\citealt{Fuerst16p13, Israel17p13}); the
observed spectrum is hardest during the peaks of the pulse cycle.

In order to model these data, we fit all three of the pulse-resolved phase bins
simultaneously and undertake a simple decomposition of these spectra into stable
and variable (\ie pulsed) components. Following the previous section, we treat the
pulsed emission simply with a \cutoffpl\ model. The `shape' parameters for this
component (\ie $\Gamma$, $E_{\rm{cut}}$) are linked across all the phase bins and
fixed to the best-fit results from the difference spectroscopy described above, but its
normalization can vary between them. We do not require the normalization of the
pulsed \cutoffpl\ component to be zero for the low-flux data, as the emission from the
accretion column can still contribute during the minimum of the pulse cycle. For the
stable emission, we return to the 2-component continuum models discussed
previously, and apply the \diskbb+\diskpbb\ combination. All the parameters for these
components, including their normalisations, are linked across all the phase bins, as
is the intrinsic neutral absorption column.

\begin{table}
  \caption{Best fit parameters obtained from our phase-resolved analysis of epoch
XN1}
\begin{center}
\begin{tabular}{c c c c c c c c c c}
\hline
\hline
\\[-0.2cm]
Component & \multicolumn{3}{c}{Parameter} \\
\\[-0.3cm]
\hline
\hline
\\[-0.2cm]
\tbabs\ & $N_{\rm H; int}$ & [$10^{20}$ cm$^{-2}$] & $8^{+2}_{-1}$ \\
\\[-0.2cm]
\diskbb\ & $kT_{\rm{in}}$ & [keV] & $0.40^{+0.01}_{-0.07}$ \\
\\[-0.2cm]
& Norm & & $1.5^{+1.4}_{-0.1}$ \\
\\[-0.2cm]
\diskpbb\ & $kT_{\rm{in}}$ & [keV] & $1.9 \pm 0.2$ \\
\\[-0.2cm]
& $p$ & & $>0.95$\tmark[a] \\
\\[-0.2cm]
& Norm & [$10^{-2}$] & $4.6^{+0.8}_{-3.6}$ \\
\\[-0.2cm]
\cutoffpl\ & $\Gamma$ & & $-0.1$\tmark[b] \\
\\[-0.2cm]
& $E_{\rm{cut}}$ & [keV] & $3.9$\tmark[b] \\
\\[-0.2cm]
& $F_{2-10}$\tmark[c] (low) & [$10^{-12}$\,\ergpcmsqps] & $1.7^{+0.2}_{-0.3}$ \\
\\[-0.2cm]
& $F_{2-10}$\tmark[c] (med) & [$10^{-12}$\,\ergpcmsqps] & $2.6^{+0.2}_{-0.3}$ \\
\\[-0.2cm]
& $F_{2-10}$\tmark[c] (high) & [$10^{-12}$\,\ergpcmsqps] & $3.6^{+0.2}_{-0.3}$ \\
\\[-0.3cm]
\hline
\\[-0.2cm]
$\chi^{2}$/DoF & & & 942/932 \\
\\[-0.3cm]
\hline
\hline
\\[-0.4cm]
\end{tabular}
\label{tab_phaseres_XN1}
\end{center}
$^a$ As before, we restrict the radial temperature index to $0.5 \leq p \leq 2.0$ \\
$^b$ These parameters have been fixed to the best-fit values from the ``pulse
on"$-$``pulse-off" difference spectroscopy (Section \ref{sec_pulsediff}). \\
$^c$ Observed fluxes for the CUTOFFPL component in the 2-10\,keV band.
\end{table}

\begin{figure}
\begin{center}
\hspace*{-0.4cm}
\rotatebox{0}{
{\includegraphics[width=240pt]{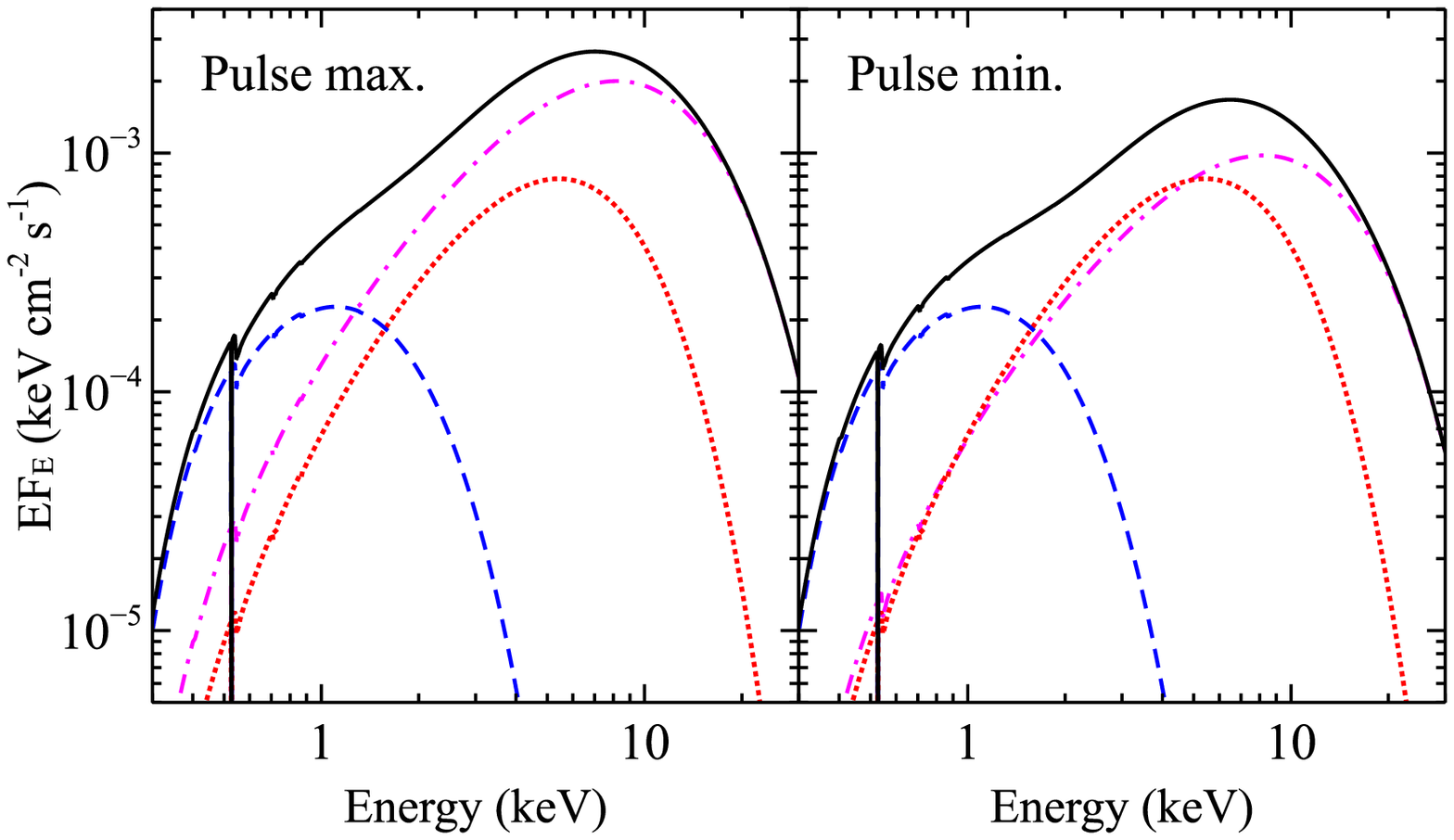}}
}
\end{center}
\caption{
The relative contributions of the various spectral components during the
peaks (\textit{left panel} and the troughs (\textit{right panel}) of the pulse cycle from
our phase-resolved analysis of epoch XN1. In both panels the total model is shown
in solid black, the DISKBB component (steady) in dashed blue, the DISKPBB
component (steady) in dotted red, and the CUTOFFPL component (pulsed) in
dash-dot magenta. The DISKBB and DISKPBB components are assumed to be
steady across the pulse cycle, and are identical in both panels.}
\label{fig_phaseres_XN1}
\end{figure}


The global fit to the phase-resolved data with this approach is excellent
($\chi^{2}$/DoF = 942/932), and the results are presented in Table
\ref{tab_phaseres_XN1}. The results for the constant \diskbb+\diskpbb\ continuum
components are broadly similar to the results obtained with the model utilizing this
combination in our time-averaged spectral analysis (see Table \ref{tab_param_av}).
This suggests that for the case of P13, the need for the \simpl\ component in the
time-averaged models is largely driven by the contribution of the pulsed emission.
The relative contribution of the \diskbb, \diskpbb\ and \cutoffpl\ components in our
analysis is shown for the peaks and the troughs of the pulse cycle in Figure
\ref{fig_phaseres_XN1}; the \cutoffpl\ model dominates the flux at the highest
energies probed by \nustar, exactly where the \simpl\ component contributes in the
time-averaged models. Indeed, the steady continuum emission does not require an
additional high-energy powerlaw tail in these fits; adding a \simpl\ component to the
\diskpbb\ component only improves the fit by $\Delta\chi^{2}$ = 6 for two additional
free parameters (although the situations are not identical, this would be below
the threshold required to claim a significant detection based on the simulations
presented above). We stress though that both the \diskbb\ and \diskpbb\ components
are required to describe the phase-resolved data; removing either significantly
degrades the global fit. Three continuum components are therefore required to fully
describe the phase-resolved data from epoch XN1: one pulsed component and two
steady components.


\section{NGC\,7793 P13: Multi-Epoch analysis}
\label{sec_ME}

\begin{figure}
\begin{center}
\hspace*{-0.6cm}
\rotatebox{0}{
{\includegraphics[width=235pt]{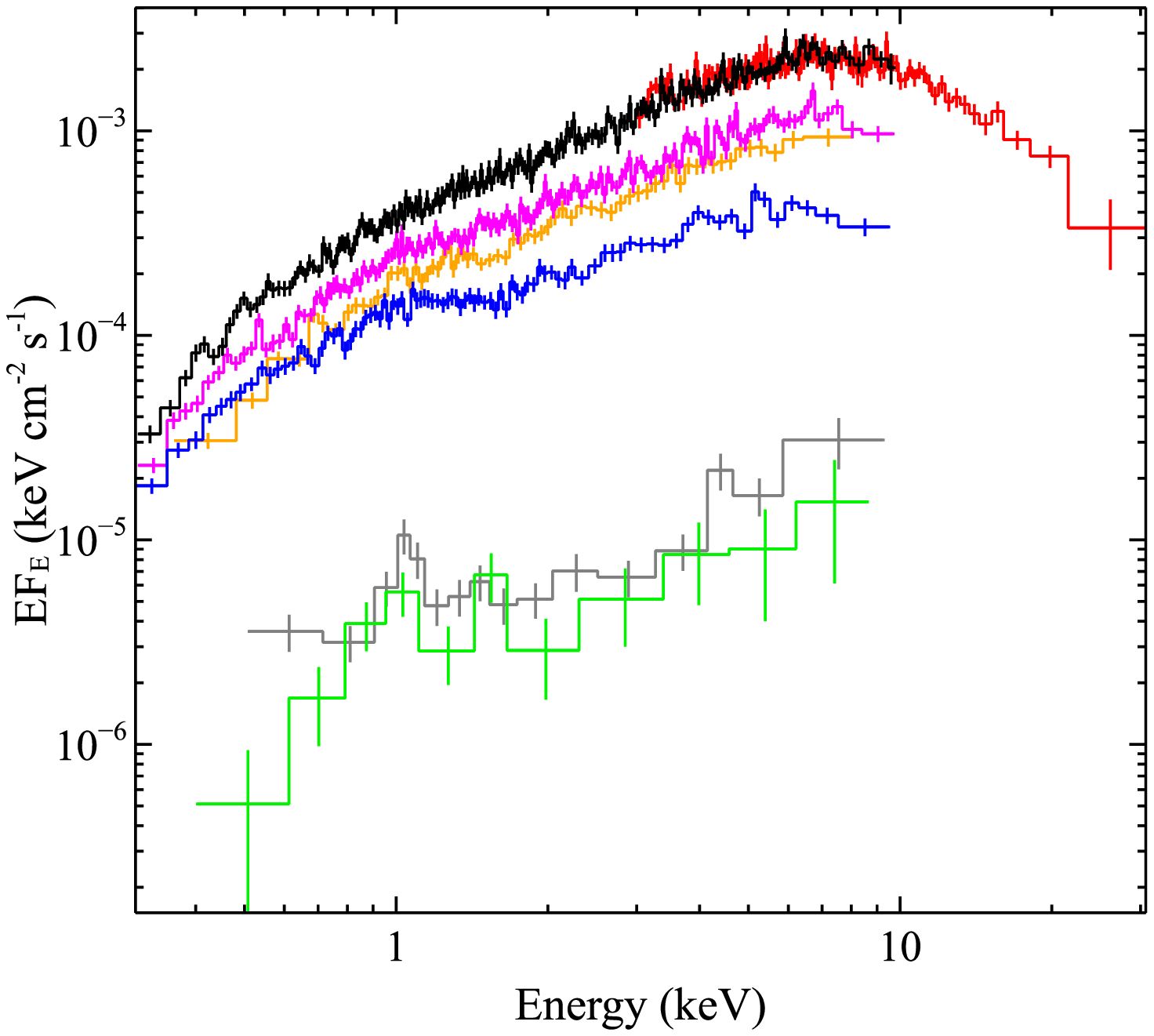}}
}
\end{center}
\caption{
A comparison of the time averaged spectra of P13 from each of the epochs
considered in this work, again unfolded through a simple model that is constant with
energy. For clarity, we only show the \xmm\ \epicpn\ and \nustar\ FPMA data. The
data from epochs XN1, X1, X2, X3, C1 and C2+C3+C4 (see Section \ref{sec_lowflux})
are shown in black (\epicpn) and red (FPMA), green, blue, magenta, orange and grey,
respectively. As before, the data have been rebinned for visual purposes.}
\label{fig_spec_ME}
\end{figure}

We next undertake a multi-epoch analysis for P13 additionally including the archival
\xmm\ datasets (see Table \ref{tab_obs}). A comparison of the average spectra from
each of these observations with the broadband \xmm+\nustar\ dataset is shown in
Figure \ref{fig_spec_ME}. The first of the archival \xmm\ observations (epoch X1) was
taken during the extended off-state that spanned mid 2011 to mid 2013
(\citealt{Motch14nat, Fuerst16p13}) and subsequently only a very low S/N spectrum is
available. However, the latter two both caught P13 at ULX luminosities, with fluxes only
a factor of $\sim$4 and $\sim$2 fainter than epoch XN1 during epochs X2 and X3,
respectively (see Table \ref{tab_pulse}). For the ULX-luminosity observations, the
0.3--10.0\,keV spectrum of P13 is clearly harder when the source is brighter, and
exhibits stronger variations at the higher energies of this bandpass. During the faintest
of these observations (epoch X2) the spectrum is most visibly doubly-peaked, as the
relative contribution of the soft thermal component is strongest during this epoch.

\begin{table*}
  \caption{Evolution in the Properties of the X-ray Pulsations across the Various
  Epochs Considered}
\begin{center}
\begin{tabular}{c c c c c c c}
\hline
\hline
\\[-0.2cm]
Epoch & $F_{\rm{av, 0.3-10}}$ & $P$ & $P_{\rm{frac, 0.5-2.0}}$ & $P_{\rm{frac, 2-10}}$ & $F_{\rm{pulse, 2-10}}$ \\
\\[-0.3cm]
& [$10^{-12}\,\ergcms$] & [ms] & [\%] & [\%] & [$10^{-12}\,\ergcms$] \\
\\[-0.3cm]
\hline
\hline
\\[-0.2cm]
X2 & $1.14 \pm 0.03$ & $419.712 \pm 0.008$ & $12 \pm 2$ & $31 \pm 3$ & $0.57 \pm 0.08$ \\
\\[-0.3cm]
X3 & $2.84 \pm 0.05$ & $418.390 \pm 0.008$ & $18 \pm 2$ & $29 \pm 2$ & $1.19 \pm 0.12$ \\
\\[-0.3cm]
XN1 & $5.19 \pm 0.07$ & $416.9513 \pm 0.0017$  & $13 \pm 1$ & $22 \pm 1$ & $1.84 \pm 0.12$ \\
\\[-0.3cm]
\hline
\hline
\\[-0.4cm]
\end{tabular}
\label{tab_pulse}
\end{center}
\end{table*}

\begin{figure}
\begin{center}
\hspace*{-0.6cm}
\rotatebox{0}{
{\includegraphics[width=235pt]{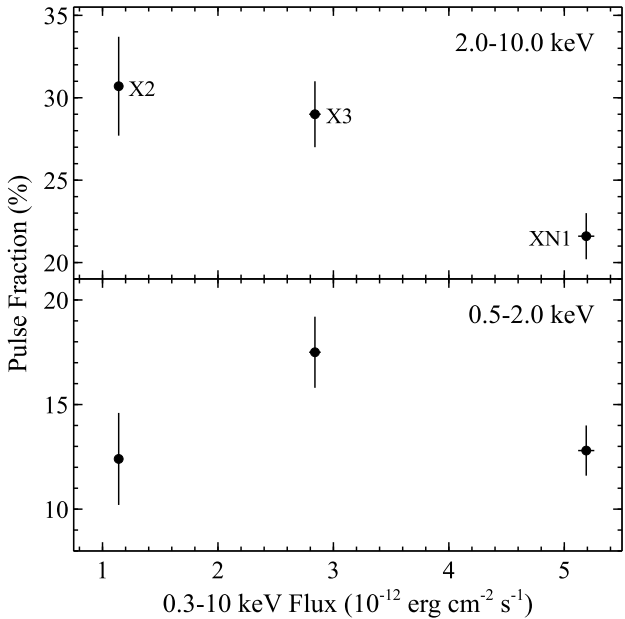}}
}
\end{center}
\caption{
Evolution of the pulse fraction in two energy bands (0.5--2.0 and
2.0--10.0\,keV) as a function of observed 0.3--10.0\,keV flux.}
\label{fig_pulsefrac_ME}
\end{figure}


\subsection{Evolution of the Pulsed Emission}
\label{sec_pulse_evol}

As noted by \cite{Fuerst16p13}, during epoch X1 P13 was too faint for any pulsations
to be detectable. In addition, none of the \chandra\ observations have sufficient
temporal resolution to see the pulsations. However, pulsations were detected in all the
subsequent epochs (X2, X3 and XN1). Table \ref{tab_pulse} summarizes the results
obtained from the pulsation searches presented in \cite{Fuerst16p13}, showing a clear
spin-up. In addition to these results, we also compute the pulse fractions for the
0.5--2.0 and 2.0--10.0\,keV bands that are common to all three epochs where
pulsations have been detected, based on the \xmm\ \epicpn\ lightcurves. Here, we
calculate the pulse fraction as (max-min)/(max+min), where max and min refer to the
average count rates at the maximum and minimum phases of the pulse cycle. The
results are shown in Figure \ref{fig_pulsefrac_ME}.

As expected, the pulse fractions are systematically higher in the harder 2.0--10.0\,keV
band. At these energies, the pulse fraction shows evidence for a secular decrease
with increasing observed flux. This suggests that there are two emission components
contributing in the 2.0--10.0\,keV band (pulsed and non-pulsed), and that the relative
contribution from the pulsed component is lower at higher fluxes. In contrast, in the
softer band the pulse fraction initially increases with flux, and then decreases again
at even higher fluxes. The evolution seen at these energies between epochs X3 and
XN1 (the two higher-flux observations) is likely related to the same effect producing
the anti-correlation in the harder band, \ie the relative contribution of the pulsed
component decreasing relative to some non-pulsed component as the flux increases.
The switch to a decreasing pulse fraction with decreasing flux is then likely related to
the increased relative importance of the cooler thermal component during epoch X2,
which is visually apparent from Figure \ref{fig_spec_ME}. This implies that the cooler
thermal component is also not pulsed, resulting in the lower soft-band pulse fractions.

\begin{figure}
\begin{center}
\hspace*{-0.6cm}
\rotatebox{0}{
{\includegraphics[width=235pt]{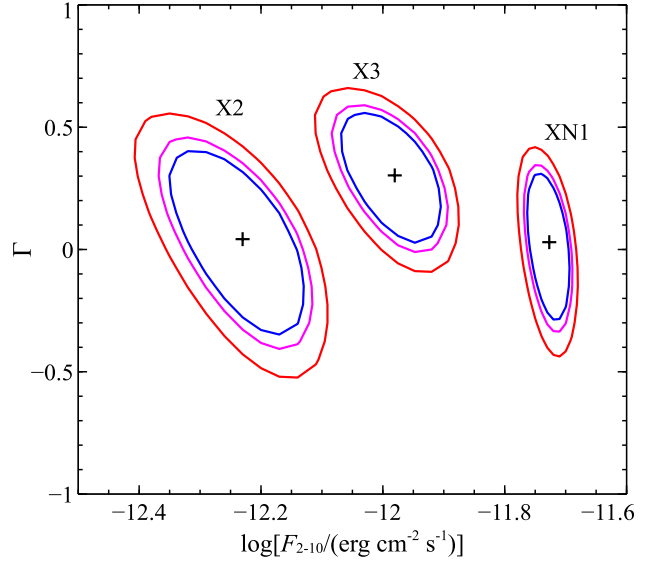}}
}
\end{center}
\caption{
2D confidence contours for $\Gamma$ and the 2--10\,keV flux for the pulsed
spectra of P13 from epochs X2, X3 and XN1. As in Figure \ref{fig_pulse_gamEc}, the
90, 95 and 99\% confidence contours for 2 parameters of interest are shown in blue,
magenta and red, respectively.}
\label{fig_ME_gamFlux}
\end{figure}


The relative evolutions seen in the pulse fractions for these two bands therefore
suggests that there are two emission components that contribute to both the hard and
the soft bands (pulsed and non-pulsed), and a third non-pulsed component that only
contributes to the soft band. This is broadly similar to the spectral decomposition
presented in Section \ref{sec_pulse}.

We also investigate the spectral evolution of the pulsed emission, by repeating the
``pulse on"$-$``pulse-off" analysis presented in Section \ref{sec_pulsediff} for the
two relevant archival \xmm\ observations. As before, here we can only make use of
the \epicpn\ data from \xmm\ for epochs X2 and X3. We fit the pulsed emission from
all three epochs (X2, X3 and XN1) simultaneously, using the \cutoffpl\ model again.
Owing to the limited bandpass available for epochs X2 and X3, we link the
high-energy cutoff between all epochs so that we can undertake a self-consistent
comparison of the slope of the pulsed emission, given the degeneracy between
$\Gamma$ and $E_{\rm{cut}}$ that can be seen in Figure \ref{fig_pulse_gamEc}.
In Figure \ref{fig_ME_gamFlux} we show the 2-D confidence contours for the photon
index and the 2--10\,keV flux for the pulsed emission from all three epochs. Although
the pulsed flux does vary, in all three cases the spectral forms are clearly consistent.
We therefore perform a final fit to the pulsed spectra in which $\Gamma$ is linked
across all the datasets as well, leaving only the normalisation of the \cutoffpl\ model
free to vary between them. With this approach, we find $\Gamma =
0.17^{+0.17}_{-0.19}$ and $E_{\rm{cut}} = 4.7^{+0.9}_{-0.7}$\,keV, consistent with
the results obtained fitting the pulsed data from epoch XN1 only. The pulsed fluxes
computed with this final analysis are given in Table \ref{tab_pulse}.

\subsection{Time-Averaged Analysis of the ULX State}
\label{sec_ME_ULX}

We now undertake a simultaneous, multi-epoch analysis of the average spectra
extracted from the epochs during which P13 exhibited ULX lumonisities. We take a
similar approach to our analysis of the phase-resolved spectra from epoch XN1 (see
Section \ref{sec_pulsephase}) and fit the data from each epoch with three continuum
components: a \cutoffpl\ model to account for the pulsed emission, and an underlying
2-component continuum to account for the non-pulsed emission, again utilizing the
\diskbb+\diskpbb\ combination. Having found that the spectral shape of the pulsed
emission remains constant throughout all the epochs from which pulsations could be
detected to date, we utilize this in our multi-epoch analysis and again fix the shape
parameters ($\Gamma$, $E_{\rm{cut}}$) of the \cutoffpl\ model to the best-fit values
obtained from our analysis of the pulsed spectra, taking the values from our global fit to
the pulsed emission from epochs X2, X3 and XN1 (see Section \ref{sec_pulse_evol}).

Given this approach, we focus here on epochs X2, X3 and XN1, since we are not able
to resolve the pulsations in the \chandra\ data from epoch C1. However, we do note
that the average spectrum from epoch C1 is very similar to epoch X3 (see Figure
\ref{fig_spec_ME}), so similar results would be expected. We fit the data from all three
epochs simultaneously in order to investigate which of the spectral parameters drive
the observed variability. This also allows us to make the simplifying assumption that
the total flux from the accretion column is proportional to the pulsed flux. As such, we
assume that the average \cutoffpl\ flux in the 2--10\,keV band (common to all epochs)
shows the same relative variations as the pulsed flux from epoch-to-epoch (see
Figure \ref{fig_ME_gamFlux} and Table \ref{tab_pulse}); the fluxes of the pulsed
emission across all three epochs are therefore controlled by a single free parameter
in our analysis, which we choose to be the pulsed flux in epoch XN1.

We also assume that the intrinsic neutral absorption column does not vary between
the epochs considered (see \eg\ \citealt{Miller13ulx}). Furthermore, during our
analysis we find that the radial temperature index for the \diskpbb\ component ($p$)
can be linked across all epochs; doing so only worsens the fit by $\Delta\chi^{2} = 1$
for 2 fewer free parameters (an F-test\footnote{See \cite{Protassov02} for
appropriate uses of the F-test.} confirms this improvement is not significant, with a
false-alarm probability of 0.6). Both of the \diskbb\ and \diskpbb\ temperatures are
found to vary. Linking either of these across all three epochs degrades the fit by
$\Delta\chi^{2} > 22$ (again for two fewer free parameters; here the F-test
gives a false-alarm probability of $< 3 \times 10^{-5}$). The global fit for the model in
which $N_{\rm{H, int}}$ and $p$ are linked across all epochs is very good, with
$\chi^{2}$/DoF = 2149/2058, and the parameter constraints are presented in Table
\ref{tab_param_ME}.

\begin{figure}
\begin{center}
\hspace*{-0.6cm}
\rotatebox{0}{
{\includegraphics[width=235pt]{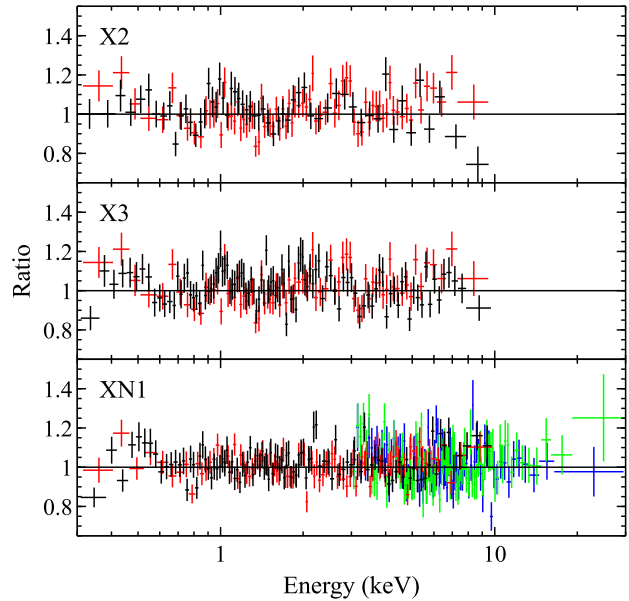}}
}
\end{center}
\caption{
Data/model ratios for our multi-epoch analysis of the time-averaged spectra
from epochs X2 (top), X3 (middle) and XN1 (bottom; see Section \ref{sec_ME_ULX}).
The ratios are plotted for the model combining DISKBB+BB+CUTOFFPL for the
intrinsic continuum; the colour-coding of the different datasets is the same as Figure
\ref{fig_spec}.}
\label{fig_ratio_ME}
\end{figure}


\begin{table*}
  \caption{Results obtained for the multi-epoch analysis of P13}
\begin{center}
\begin{tabular}{c c c c c c c}
\hline
\hline
\\[-0.2cm]
Model & \multicolumn{2}{c}{Parameter} & Global & \multicolumn{3}{c}{Epoch} \\
\\[-0.3cm]
Component & & & & X2 & X3 & XN1 \\
\\[-0.3cm]
\hline
\hline
\\[-0.2cm]
\multicolumn{7}{c}{Continuum: DISKBB+DISKPBB+CUTOFFPL} \\
\\[-0.2cm]
\tbabs\ & $N_{\rm{H,int}}$ & [$10^{20}$ cm$^{-2}$] & $7.7 \pm 0.8$ \\
\\[-0.3cm]
\diskbb & $T_{\rm{in}}$ & [keV] & & $0.34 \pm 0.01$ & $0.417^{+0.008}_{-0.026}$ & $0.421^{+0.009}_{-0.022}$ \\
\\[-0.3cm]
& Norm & & & $1.24^{+0.32}_{-0.10}$ & $0.85^{+0.25}_{-0.07}$ & $1.25^{+0.25}_{-0.11}$ \\
\\[-0.3cm]
\diskpbb\ & $T_{\rm{in}}$ & [keV] & & $1.70 \pm 0.15$ & $2.14^{+0.14}_{-0.11}$ & $2.15^{+0.10}_{-0.06}$ \\
\\[-0.3cm]
& $p$ & & $>1.52$\tmark[a] \\
\\[-0.3cm]
& Norm & [$10^{-2}$] & & $1.3^{+0.4}_{-0.5}$ & $1.9^{+0.3}_{-0.7}$ & $4.0^{+0.1}_{-1.4}$ \\
\\[-0.3cm]
\cutoffpl\ & $\Gamma$ & & $0.17$\tmark[b] \\
\\[-0.3cm]
& $E_{\rm{cut}}$ & [keV] & $4.7$\tmark[b] \\
\\[-0.3cm]
& $F_{2-10}$ & [$10^{-13}$\,\ergpcmsqps] & & $5.5$\tmark[c] & $11.3$\tmark[c] & $17.6^{+2.0}_{-2.8}$ \\
\\[-0.3cm]
\hline
\\[-0.2cm]
\chisq/DoF & & & 2149/2058 \\
\\[-0.3cm]
\hline
\hline
\\[-0.2cm]
\multicolumn{7}{c}{Continuum: DISKBB+BB+CUTOFFPL} \\
\\[-0.2cm]
\tbabs\ & $N_{\rm{H,int}}$ & [$10^{20}$ cm$^{-2}$] & $7.1^{+0.8}_{-0.7}$ \\
\\[-0.3cm]
\diskbb & $T_{\rm{in}}$ & [keV] & & $0.35 \pm 0.02$ & $0.45 \pm 0.03$ & $0.48 \pm 0.03$ \\
\\[-0.3cm]
& Norm & & & $1.05^{+0.27}_{-0.21}$ & $0.63^{+0.17}_{-0.13}$ & $0.80^{+0.19}_{-0.15}$ \\
\\[-0.3cm]
\bb\ & $T_{\rm{in}}$ & [keV] & & $1.1 \pm 0.1$ & $1.5 \pm 0.1$ & $1.53 \pm 0.04$ \\
\\[-0.3cm]
& Norm & [$10^{-6}$] & & $2.4 \pm 0.6$ & $11.1 \pm 1.5$ & $27.3^{+2.4}_{-2.2}$ \\
\\[-0.3cm]
\cutoffpl\ & $\Gamma$ & & $0.17$\tmark[b] \\
\\[-0.3cm]
& $E_{\rm{cut}}$ & [keV] & $4.7$\tmark[b] \\
\\[-0.3cm]
& $F_{2-10}$ & [$10^{-13}$\,\ergpcmsqps] & & $6.9$\tmark[c] & $14.4$\tmark[c] & $22.3^{+1.4}_{-1.6}$ \\
\\[-0.3cm]
\hline
\\[-0.2cm]
\chisq/DoF & & & 2144/2059 \\
\\[-0.3cm]
\hline
\hline
\\[-0.4cm]
\end{tabular}
\label{tab_param_ME}
\end{center}
\flushleft
$^a$ As before, we restrict the radial temperature index to $0.5 \leq p \leq 2.0$ \\
$^b$ These parameters have been fixed to the best-fit values from the global fit to the
multi-epoch ``pulse on"$-$``pulse-off" difference spectra (Section \ref{sec_pulse_evol}). \\
$^c$ The CUTOFFPL fluxes for epochs X2 and X3 are scaled relative to epoch XN1
based on the relative pulsed fluxes seen from these epochs (see Table \ref{tab_pulse}). \\
\end{table*}

\begin{figure*}
\begin{center}
\hspace*{-0.6cm}
\rotatebox{0}{
{\includegraphics[width=470pt]{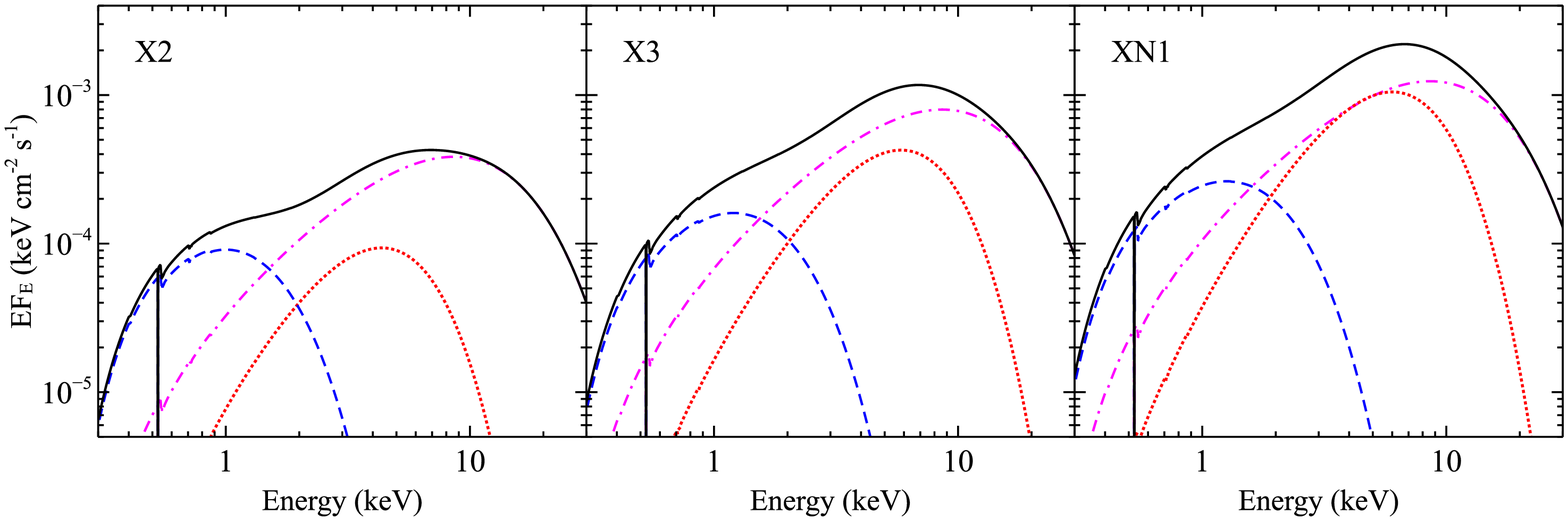}}
}
\end{center}
\caption{
The relative contributions of the various spectral components from our
multi-epoch analysis of the time-averaged spectra from epochs X2 (left), X3 (middle)
and XN1 (right). In all panels the total model is shown in solid black, the DISKBB
component (steady) in dashed blue, the BB component (steady) in dotted red, and the
CUTOFFPL component (pulsed) in dash-dot magenta, respectively.}
\label{fig_eem_ME}
\end{figure*}


One notable result common to all our spectral analyses is that the radial temperature
index for the \diskpbb\ component is rather steep; here we find $p > 1.52$ (see also
Sections \ref{sec_avspec} and \ref{sec_pulsephase}). This implies that the emission
from this component is significantly more peaked, i.e. dominated by a narrow range of
temperatures, than a standard thin disk continuum (which should give $p = 0.75$;
\citealt{Shakura73}). We therefore test here whether a range of temperatures is really
required for this component, and replace \diskpbb\ with a single blackbody (\bb)
component. The fit in this case is actually slightly improved over the \diskpbb\ model,
despite having one fewer free parameter: $\chi^{2}$/DoF = 2144/2059 (this is likely due
to the upper limit of 2.0 we place on $p$ in our previous models). In contrast, the lower
temperature component does require a range of temperatures; replacing the \diskbb\
component with a single \bb\ worsens the fit by $\Delta\chi^{2} > 30$ (for the same
number of free parameters) regardless of the model used for the hotter component
(\diskpbb\ or \bb).

The results for the model in which the hotter component is treated as a single \bb\ are
also given in Table \ref{tab_param_ME}, and we show the data/model ratios for each
epoch in Figure \ref{fig_ratio_ME}. We also show the relative contributions of the
various components for the best-fit model in Figure \ref{fig_eem_ME}; the evolution of
the various components is qualitatively consistent with that expected from the
evolution in the soft and hard band pulse fractions (Figure \ref{fig_pulsefrac_ME}).
Lastly, as a further sanity check, we compare the total flux for the \cutoffpl\ model in
epoch XN1 (against which the other epochs are scaled) obtained with this model to
that expected based on the phase-resolved analysis of this epoch (Section
\ref{sec_pulsephase}) and the pulsed flux observed (Table \ref{tab_pulse}). The flux for
the \cutoffpl\ component obtained for the minimum of the pulse cycle is $\sim$1.7
$\times 10^{-12}$ \ergpcmsqps\ in the 2--10\,keV band (see Table
\ref{tab_phaseres_XN1}). For a perfectly symmetric sinusoidal pulse profile, the total
\cutoffpl\ flux would be expected to be the sum of this flux and half of the pulsed flux, \ie
$\sim$2.6 $\times 10^{-12}$\,\ergpcmsqps. This is very similar to the total \cutoffpl\ flux
obtained for epoch XN1 in our multi-epoch spectral analysis with the
\diskbb+\bb+\cutoffpl\ model (Table \ref{tab_param_ME}), particularly when accounting
for statistical uncertainties and the fact that the pulse profile is not perfectly sinusoidal
(\citealt{Fuerst16p13}), suggesting that our spectral decomposition in this case is
internally self-consistent.

\subsection{The Low-Flux Observations}
\label{sec_lowflux}

\begin{figure}
\begin{center}
\hspace*{-0.6cm}
\rotatebox{0}{
{\includegraphics[width=235pt]{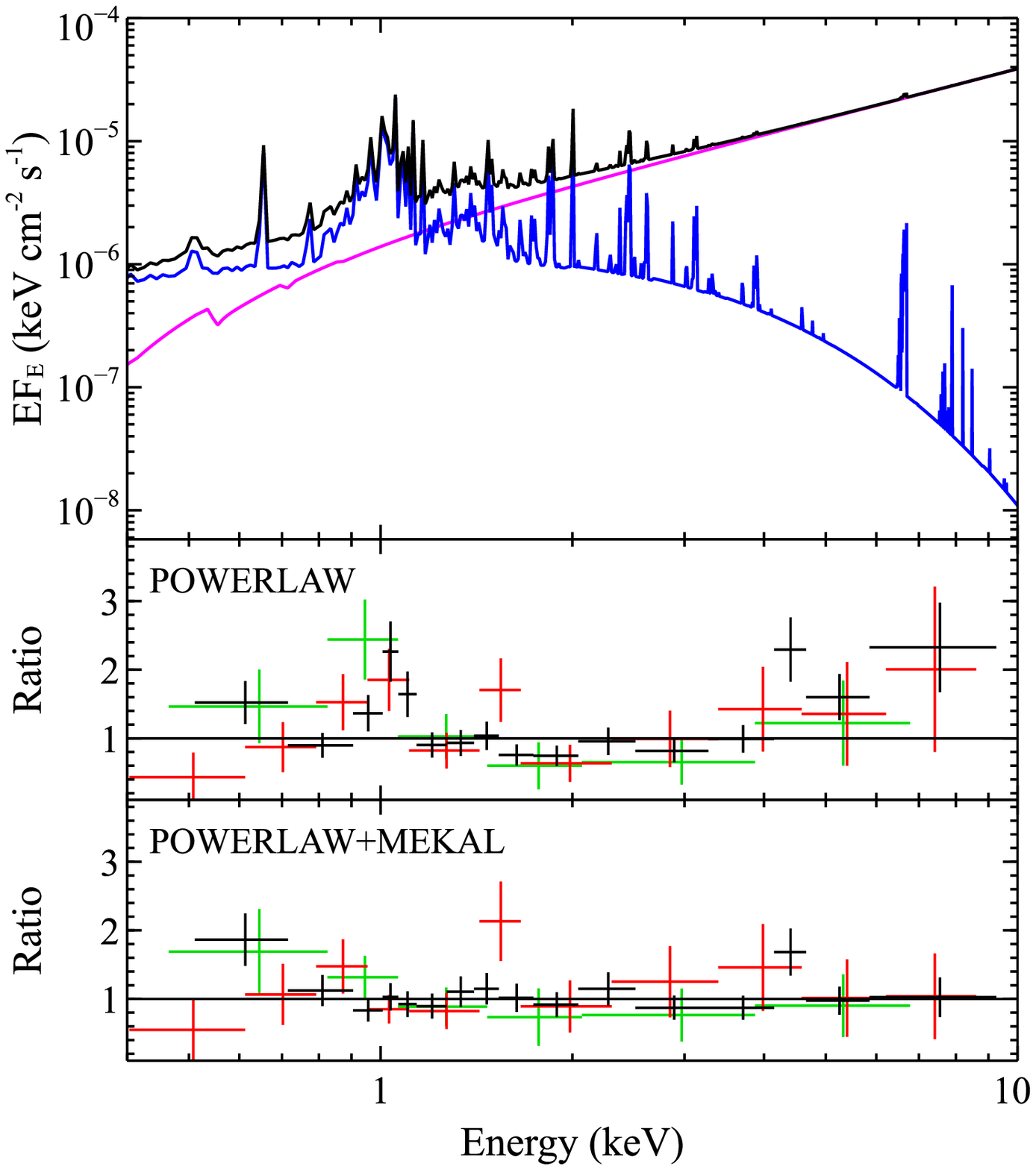}}
}
\end{center}
\caption{
\textit{Top panel:} The best-fit model for the integrated \chandra\ data
obtained during the off-state (\ie epochs C2+C3+C4; see Section \ref{sec_lowflux}).
The total model, and the relative contributions of the powerlaw and the \mekal\
components are shown in black, magenta and blue, respectively. \textit{Middle panel:}
data/model ratio for the powerlaw-only model, fit to both the \chandra\ and \xmm\
off-state datasets. Clear residuals remain, particularly at $\sim$1\,keV. \textit{Bottom
panel:} same as the middle panel, but for the powerlaw $+$ \mekal\ model. In both
the middle and bottom panels, the \chandra\ ACIS data and the \xmm\ \epicpn\ and
\epicmos\ data are shown in black, red and green, respectively.}
\label{fig_lowflux}
\end{figure}

In addition to the ULX luminosity observations, we also undertake a simple analysis
of the average spectra extracted from the observations taken during the off-state
(epochs C2, C3, C4 and X1). Although there are some flux variations, the spectra from
these epochs are all broadly consistent (key spectral parameters agree within their
90\% errors when fit individually), so to maximise the S/N we co-add the \chandra\ data
to form a single spectrum using \addascaspec. The combined \chandra\ spectrum from
epochs C2+C3+C4 and the \epicpn\ data from epoch X1 are shown in Figure
\ref{fig_spec_ME}. In order to get the best constraints for this state, we fit these
\chandra\ and \xmm\ datasets simultaneously with some simple models. We first apply
a simple absorbed powerlaw continuum, fixing the intrinsic neutral absorption to
$N_{\rm{H; int}} = 8 \times 10^{20}$\,\pcmsq\ based on the fits to the broadband data,
and assuming a common photon index for the \chandra\ and \xmm\ datasets. However,
we find that this does not fit the data ($\chi^{2}$/DoF = 63/29) and leaves obvious
residuals. These are particularly prominent around $\sim$1\,keV (see Figure
\ref{fig_lowflux}), and are reminiscent of the contribution from thermal plasma emission
at low energies.

\begin{table}
  \caption{Best fit parameters obtained for the off-state datasets}
\begin{center}
\begin{tabular}{c c c c c c}
\hline
\hline
\\[-0.2cm]
Parameter & & Global & \chandra\ & \textit{XMM} \\
\\[-0.3cm]
\hline
\hline
\\[-0.2cm]
$kT_{\rm{mekal}}$ & [keV] & $1.3^{+0.3}_{-0.2}$ \\
\\[-0.3cm]
Norm & [$10^{-6}$] & $5 \pm 2$ \\
\\[-0.3cm]
$\Gamma$ & & $0.7^{+0.4}_{-0.5}$ \\
\\[-0.3cm]
Norm & [$10^{-6}$] & & $1.8^{+1.2}_{-0.9}$ & $0.8^{+0.9}_{-0.5}$ \\
\\[-0.3cm]
\hline
\\[-0.175cm]
$\chi^{2}$/DoF & & 28/27 \\
\\[-0.3cm]
\hline
\\[-0.2cm]
$L_{\rm{tot}}$\tmark[a] & \multirow{2}{*}{[$10^{37}$\,\ergps]} & & $7.9^{+1.2}_{-1.3}$ & $4.5^{+1.4}_{-1.6}$ &  \\
\\[-0.3cm]
$L_{\rm{po}}$\tmark[a] & & & $6.5 \pm 1.4$ & $3.0^{+1.2}_{-1.7}$ \\
\\[-0.3cm]
\hline
\hline
\\[-0.4cm]
\end{tabular}
\label{tab_lowflux}
\end{center}
$^{a}$ Observed luminosities in the 0.3--10.0\,keV band for the total emission and the
powerlaw component, respectively, assuming isotropic emission
\end{table}

We therefore add a \mekal\ plasma component to the model. We assume this is only
absorbed by Galactic absorption, that this component has solar abundances, and
that the temperature remains constant between the \chandra\ and \xmm\ datasets, but
we also find that the normalisations for this component are consistent between the
datasets, and therefore also link this parameter in our final model. The only parameter
that is allowed to vary between the \xmm\ and \chandra\ spectra is the normalisation of
the powerlaw component. This model fits the off-state data very well ($\chi^{2}$/DoF =
28/27), and resolves the $\sim$1\,keV residuals. We stress that both the \mekal\ and
the powerlaw components are required to fit the data; removing the latter results in a
significantly worse fit again ($\chi^{2}$/DoF = 61/29). The best-fit model for the
\chandra\ spectrum is shown in Figure \ref{fig_lowflux}, and the parameter constraints
are given in Table \ref{tab_lowflux}.

The \mekal\ component contributes at low-energies, fitting the $\sim$1\,keV feature
with iron L emission, but in addition to this a very hard powerlaw continuum ($\Gamma
= 0.7^{+0.4}_{-0.5}$) is required at higher energies, which we assume to be related to
residual accretion onto the neutron star. However, the fact that the normalizations for
the \mekal\ component are consistent for the \xmm\ and \chandra\ datasets, despite
their significantly different extraction regions, suggests that this is not faint, extended
emission from the broader galaxy, but is likely also be associated with P13. However,
even if this is the case it is not clear that this emission is directly powered by accretion,
so we compute the observed fluxes for both the total emission and just the powerlaw
component (see Table \ref{tab_lowflux}). The latter provides the majority of the
observed flux in the 0.3--10.0\,keV band in both the \chandra\ and the \xmm\ datasets;
the residual accretion power is orders of magnitude lower than the peak flux observed
from epoch XN1, regardless of the power source for the plasma emission.

\begin{figure*}
\begin{center}
\rotatebox{0}{
{\includegraphics[width=490pt]{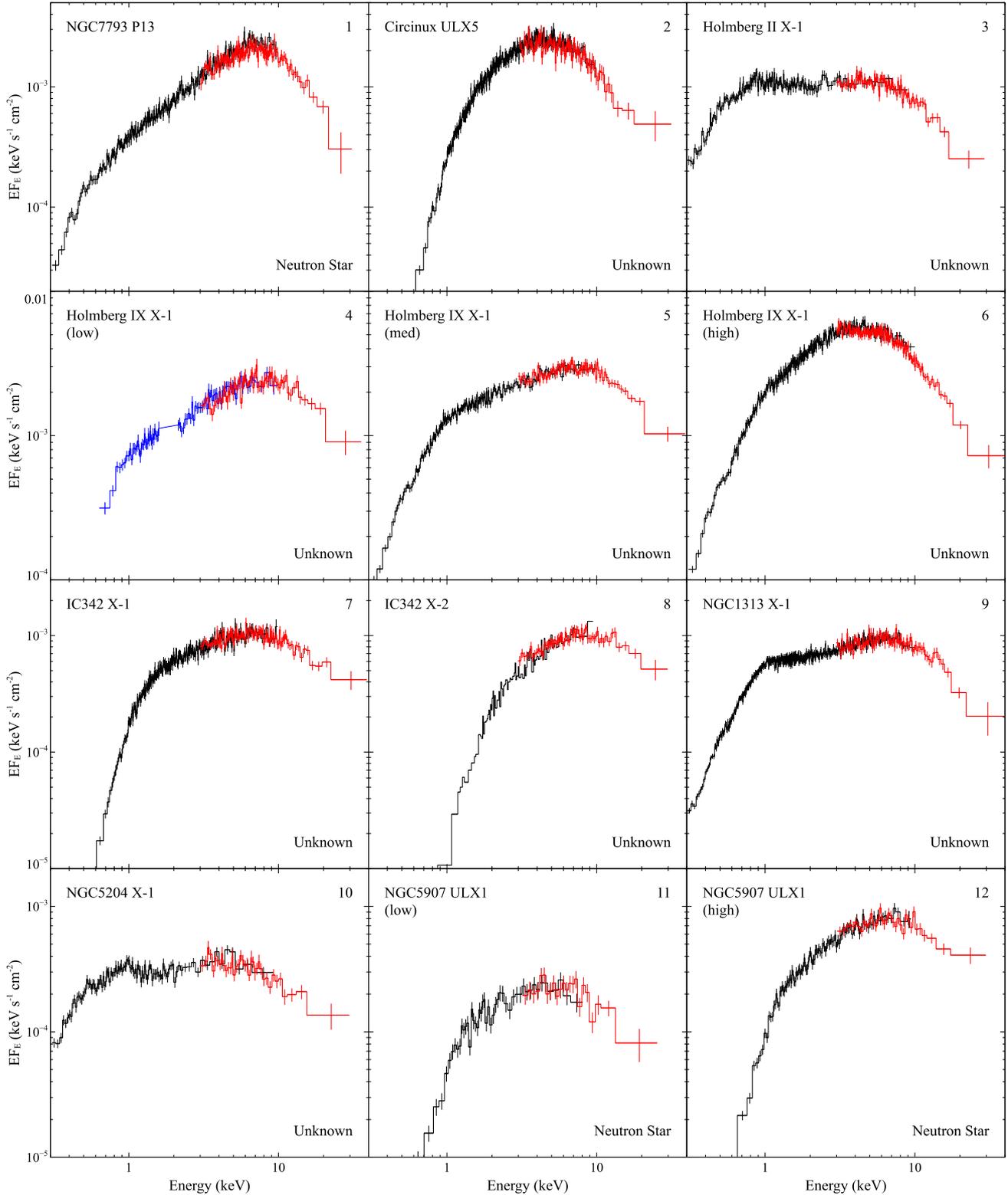}}
}
\end{center}
\caption{
Broadband spectra of the ULX population with simultaneous coverage in
both soft (\xmm, \suzaku) and hard (\nustar) X-rays; as for Figure \ref{fig_spec}
(\textit{left}), the data have been unfolded through a model simply consisting of a
constant. Only sources which suffer from negligible source confusion and which are
robustly detected out to at least 20\,keV are included, and for clarity, only data from
the \xmm\ EPIC-pn (black), \nustar\ FPMA (red) and \suzaku\ front-illuminated XIS
(blue) detectors are shown (the only exception is NGC\,5907 ULX, where for \nustar\
we show the combined FPMA+FPMB data owing to the lower S/N in this case). The
sample includes P13 (panel 1; this work), Circinus ULX5 (panel 2; adapted from
\citealt{Walton13culx}), Holmberg II X-1 (panel 3; adapted from \citealt{Walton15hoII}),
Holmberg IX X-1 (panels 4--6; adapted from \citealt{Walton17hoIX}), IC\,342 X-1 and 
X-2 (panels 7 and 8, respectively; adapted from \citealt{Rana15}), NGC\,1313 X-1
(panel 9; adapted from \citealt{Bachetti13}), NGC\,5204 X-1 (panel 10; adapted from
\citealt{Mukherjee15}) and NGC\,5907 ULX (panels 11--12; adapted from
\citealt{Fuerst17ngc5907}).}
\label{fig_sample}
\end{figure*}


\section{The Broadband ULX Sample}
\label{sec_ulxpop}

For comparison with \p13, we also compiled the sample of ULXs with broadband
spectra available in the literature at the time of writing. We focus only on sources for
which \nustar\ has provided a robust detection out at least 20keV, which are not
significantly confused with any other X-ray sources, and for which simultaneous data
from at least one other mission with high S/N soft X-ray coverage (\xmm\ and/or
\suzaku) exist. This sample includes: Circinus ULX5 (\citealt{Walton13culx}),
Holmberg II X-1 (\citealt{Walton15hoII}), Holmberg IX X-1 (\citealt{Walton14hoIX,
Walton17hoIX, Luangtip16}), IC\,342 X-1 and X-2 (\citealt{Rana15}), NGC\,1313 X-1
(\citealt{Bachetti13, Miller14, Walton16ufo}), NGC\,5204 X-1 (\citealt{Mukherjee15})
and NGC\,5907 ULX (\citealt{Walton15, Israel17, Fuerst17ngc5907}). Where
necessary, we reprocessed the data with the latest calibrations, largely following the
reduction procedures outlined in the referenced works, and also included the \nustar\
`spacecraft science' data (see Section \ref{sec_red}). Notable exclusions are
NGC\,1313 X-2, which is not detected much above 10\,keV (\citealt{Bachetti13}), and
M82 X-1 and X-2, which are strongly blended in the \nustar\ data (\eg\
\citealt{Bachetti14nat, Brightman16m82a, Brightman16m82b}). We also do not
consider NGC\,5643 X-1 or M51\,X-8; both are only weakly detected by \nustar\ above
10\,keV in short observations, and neither of the \nustar\ observations of these sources
have simultaneous high S/N soft X-ray coverage (\citealt{Annuar15, Krivonos16,
Earnshaw16}). Finally, we also note that there is a further \nustar\ observation of
IC\,342 that we also do not include, as again this has no simultaneous high S/N soft
X-ray coverage (\citealt{Shidatsu17}). The majority of the sources considered here are
also included in the ULX sample recently presented by \cite{Pintore17}.

The time-averaged broadband spectra for the selected sample, including \p13, are
shown in Figure \ref{fig_sample} (in the case of Holmberg IX X-1, we only show three
of the six broadband observations available, corresponding to epochs 1, 2 and 5 as
defined in \citealt{Walton17hoIX}; the remaining three are similar to either
the low or medium-flux states shown). There is clearly some diversity among the
observed spectra; the lowest energies in particular are influenced by different levels of
neutral absorption. However, there are also many similarities. For all the sources with
low absorption, evidence for two thermal continuum components can be seen in the
spectra below 10\,keV (although the relative contributions of these components can
vary both between sources and over time for an individual source, also helping to
increase the diversity of the observed spectral shapes), similar to P13.
\cite{Motch14nat} also note that the spectrum of P13 below 10\,keV is fairly
representative of the broader ULX population. The qualitative similarity of the \nustar\
data in all cases is particularly striking; all the high-energy spectra break in a
fairly narrow energy band above $\sim$3\,keV to a steep continuum above 10\,keV.
Evidence for an additional high-energy powerlaw tail (or at least, an additional
high-energy continuum component) has now been seen in several other systems in
addition to P13 (Circinus ULX5: \citealt{Walton13culx}; Holmberg II X-1:
\citealt{Walton15hoII}; Holmberg IX X-1: \citealt{Walton14hoIX, Walton17hoIX};
NGC\,5204 X-1: \citealt{Mukherjee15}; NGC\,5907 ULX: \citealt{Fuerst17ngc5907}).
Despite the clear diversity among the spectra, the basic spectral components
present in the broader ULX population, for which the nature of the accretors generally
remains unknown, therefore appear similar to \p13\ (see also \citealt{Pintore17}).

\section{Discussion}
\label{sec_dis}

We have performed a detailed spectral analysis of the ultraluminous X-ray source
\p13, focusing on the first high-energy detection of the source by \nustar. The recent
detection of X-ray pulsations (\citealt{Fuerst16p13, Israel17p13}) firmly identify it as a
neutron star accretor, meaning the peak luminosity observed ($L_{\rm{X, peak}} \sim
10^{40}$\,\ergps) is highly super-Eddington. The low level of absorption towards P13
and the lack of source confusion makes it a key laboratory for understanding the
super-Eddington accretion onto neutron stars, and also their relation to the broader
ULX population. Through a combination of time-averaged, phase-resolved and
multi-epoch analyses, we find that three spectral components are required to fit the
data: two thermal blackbody components that contribute below $\sim$10\,keV (with
temperatures of the hotter and cooler components varying from $\sim$1--1.5\,keV and
from $\sim$0.3--0.5\,keV between the observed epochs, respectively), and a third
component that extends to higher energies. As discussed in Section \ref{sec_ulxpop},
these components are qualitatively similar to those seen in other ULXs with broadband
coverage.

\subsection{The Thermal Components}
\label{sec_BB}

Here, we examine the behaviour of these thermal components in order to try and
determine their origin. One subtle difference between P13 and the broader ULX
population worth noting is that for the hotter of the two components we find that the
data prefer a steep radial temperature index (\ie $p \gtrsim 1.0$) when fit with a disk
model (\diskpbb). This is true for all our various analyses of the P13 data (the
time-averaged and phase-resolved analysis of epoch XN1, and the multi-epoch
analysis of epochs X2, X3 and XN1). In fact, for our multi-epoch analysis, the fit is
slightly improved if we do not allow for a range of temperatures at all, and fit the
hotter component as a single-temperature blackbody (Section \ref{sec_ME_ULX}).
For other ULX systems, the radial temperature indices obtained for the hotter
component have typically been much flatter, with $p < 0.75$ (\eg\
\citealt{Walton15hoII, Walton17hoIX}).

\begin{figure}
\begin{center}
\hspace*{-0.6cm}
\rotatebox{0}{
{\includegraphics[width=235pt]{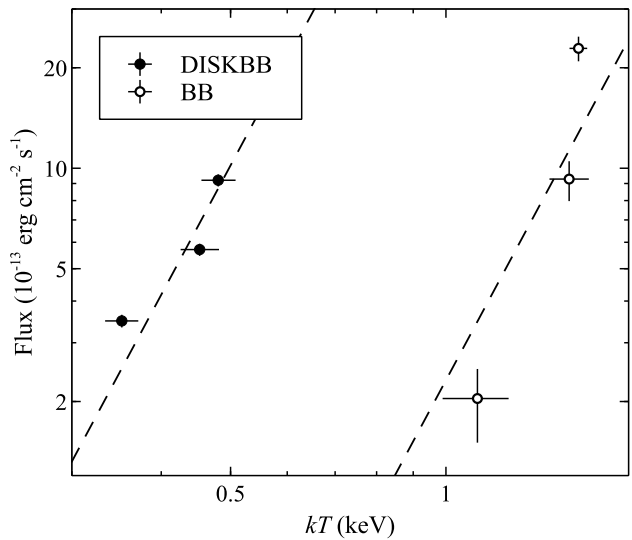}}
}
\end{center}
\caption{
Flux vs temperature for both the \diskbb\ (filled circles) and \bb\ (empty
circles) components in our multi-epoch spectral analysis. For illustration, we also
show example $F \propto T^{4}$ relations (dashed lines).}
\label{fig_LT}
\end{figure}

In Figure \ref{fig_LT} we plot the flux of each of the \diskbb\ and \bb\ components
(computed over 0.01--100\,keV, a sufficiently broad band to effectively be bolometric)
against their temperatures. With only three data points to date, we do not formally fit any
relations to these data, but for comparison we do overlay illustrative $F \propto T^{4}$
relations as expected for blackbody emission from a constant emitting area. The lower
energy \diskbb\ component does appear to follow $F \propto T^{4}$ reasonably well,
suggesting a roughly constant emitting area, but the same is not true for the \bb\
component. In the latter case, epoch XN1 sits significantly off any extrapolation of an $F
\propto T^{4}$ relation from the lower flux data, suggesting the emitting area for this
component may be variable.

For the blackbody component, the maximum and minimum flux/temperature
combinations (from epochs XN1 and X2, respectively) imply emitting areas with radii
of $\sim$40--70\,km from standard blackbody theory, or equivalently $\sim$20--35\,\rg\
for a 1.4\,\msun\ neutron star (where \rg\ = $GM/c^{2}$ is the gravitational radius). This
is comfortably larger than typical radii for the entire neutron star ($\sim$10\,km),
suggesting that this component must come from the accretion flow rather than from the
surface of the neutron star itself, particularly given that the emission from the surface of
a magnetised neutron star should primarily arise from small hotspots around the
magnetic poles onto which the accretion is channelled. For the lower temperature
\diskbb\ component, the fact that this component is required to have a range of
temperatures already suggests that this must also be associated with the accretion flow.
This is further supported by the emitting radius implied by the \diskbb\ normalizations,
which are proportional to $R_{\rm{in}}^{2}\cos\theta/f_{\rm{col}}^{4}$. Here, $R_{\rm{in}}$
and $\theta$ are the inner radius and the inclination of the disc, and $f_{\rm{col}}$ is the
colour correction factor relating the observed `colour' temperature to the effective
temperature at the midplane of the disc ($f_{\rm{col}} = T_{\rm{col}}/T_{\rm{eff}}$). This
latter quantity provides a simple empirical correction accounting for the complex physics
in the disk atmosphere. Even assuming that both $\cos\theta$ and $f_{\rm{col}}$ are
unity (\ie no colour correction and a face-on disc; note that this is also implicitly assumed
for the hotter thermal component in our use of standard blackbody theory above), which
would give the smallest radii, we find that $R_{\rm{in}} \sim 280$-360\,km
($\sim$140--180\,\rg). This is at least an order of magnitude larger than typical neutron
star radii. We will re-visit these radii in Section \ref{sec_NS_flow}.

\subsection{The Pulsed Emission/High-Energy Tail}

In addition to the two thermal components, our analysis of the time-averaged
spectrum finds that a third continuum component is required above 10\,keV, as all
two-component thermal models leave a clear excess at the highest energies probed
by \nustar\ (Figure \ref{fig_spec}). As discussed previously, similar high-energy
excesses have been seen in the average spectra of several other ULX systems now.
In these previous works, the excess has typically been modelled as steep ($\Gamma
\sim 3$) high-energy powerlaw tail to the lower-energy thermal continuum
components, assumed to arise through Compton up-scattering similar to the X-ray
coronae seen in sub-Eddington systems. When considering the time-averaged
spectrum for P13, the same model fits the data well.

However, through phase-resolved analyses of the P13 data, we find that this excess
is actually associated with the pulsed emission component arising from the magnetically
channelled accretion columns. rather than from a traditional X-ray corona. The emission
from the accretion column dominates over the two thermal components above 10\,keV,
naturally explaining the fact that the pulsed fraction increases with increasing energy
below 10\,keV, and appears to saturate at higher energies (\citealt{Fuerst16p13}). This
emission is well described with a simple \cutoffpl\ model, which has a very hard rise
($\Gamma \sim 0.15$) below 10\,keV before breaking to a steep spectrum above
10\,keV thanks to a low cutoff energy ($E_{\rm{cut}} \sim 4.5$\,keV). This is qualitatively
similar to the pulsed emission from the the first ULX pulsar discovered, M82 X-2
(\citealt{Brightman16m82a}, although for M82 X-2 the rise is not quite as hard and the
cutoff at a higher energy, see Figure \ref{fig_pulse_gamEc}). The average spectrum of
X-2 cannot easily be disentangled from M82 X-1, complicating spectral analyses.
Nevertheless, this does show that the emission from the accretion column also extends
up to these energies, and the pulsed fraction for M82 X-2 is also known to increase with
increasing energy (\citealt{Bachetti14nat}), despite the confusion with X-1. We therefore
expect that the emission from the accretion column also dominates at the highest
energies in this system.

\subsection{Super-Eddington Accretion onto Pulsar ULXs}
\label{sec_NS_flow}

For super-Eddington accretion onto a compact object, the inner regions of the
accretion flow are expected to have a large scale-height, owing to the increased
support by the intense radiation pressure (\eg\ \citealt{Shakura73, Abram88,
Poutanen07, Dotan11}). This occurs interior to the point at which the accretion flow
is locally at the Eddington limit, a radius commonly referred to as the `spherization
radius' (\rsp). The geometry of a super-Eddington accretion flow is expected to
resemble a funnel-like structure close to the central accretor that transitions to a more
standard thin disk at larger radii. The central accretors powering ULXs have generally
been assumed to be black holes in the literature, and theoretical effort has therefore
primarily focused on super-Eddington accretion onto these objects, but \cite{King08}
argue that the same framework should hold for super-Eddington accretion onto a
neutron star (at least, in regions of the disk not significantly influenced by the neutron
star's magnetic field). 

In the context of this model, the two thermal components required to successfully
reproduce ULX spectra below 10\,keV are often interpreted as arising from different
regions in a super-Eddington accretion flow, with the hotter and cooler components
arising from inside and outside the inner funnel, respectively (\eg\ \citealt{Middleton15,
Luangtip16}, but see \citealt{Miller14} for an alternative scenario). Large scale-height,
optically-thick accretion flows are expected to show $p < 0.75$ owing to the significant
photon advection that should occur in this region (\citealt{Abram88}), as is often seen
for the hotter component in ULX spectral modelling (\eg\ \citealt{Walton15hoII,
Walton17hoIX}), but is not seen for P13.

For ULX pulsars, the influence of the neutron star's magnetic field also needs to be
considered, This will disrupt the disk at the magnetospheric radius (\rmag), at which
point the accreting material is forced to follow the field lines down onto the neutron star.
Since the discovery of the first ULX pulsar (\citealt{Bachetti14nat}), an extremely wide
range of magnetic fields have been proposed in the literature in order to explain the
observed properties of these sources, from $B \sim 10^{9-14}$\,G (\eg\
\citealt{Kluzniak15, Mushtukov15}). This naturally leads to a range of predictions for the
accretion geometry, depending on the relative sizes of \rsp\ and \rmag. If \rmag\ $>$ \rsp\
(high B-field), then the magnetic pressure will terminate the accretion flow before the
large scale-height inner funnel can form. Alternatively, if \rmag\ $\ll$ \rsp\ (low B-field),
then the disk will only truncate very close to the neutron star, allowing the inner funnel to
form unhindered, and the geometry will resemble that described above. Finally, for more
intermediate configurations in which \rmag\ $\lesssim$ \rsp\ (\ie smaller, but still roughly
similar), then the inner funnel should begin to form, but will be terminated prematurely by
the magnetic field.

The fact that even after accounting for the pulsed emission from the accretion
column we still require two thermal components to reproduce the P13 data supports the
idea that there are two distinct regions to the accretion flow beyond \rmag, suggesting
that \rmag\ $<$ \rsp. One might therefore expect to be able to associate the
characteristic radii of the hotter and cooler components with \rmag\ and \rsp, respectively.
Our previous crude estimates of these radii are qualitatively consistent with this scenario,
with the hotter component arising from a smaller region. Interestingly, the radii of these
two components appear to be roughly comparable (differing by a factor of $\sim$6;
Section \ref{sec_BB}).

Several additional factors need to be considered to assess whether these radii
could really be similar in reality. The values presented in Section \ref{sec_BB} were
estimated by setting both $f_{\rm{col}}$ and $\cos\theta$ to unity in order to assess
whether either component could be associated with the neutron star surface. Both of
these quantities can have a strong influence on the absolute radii inferred. However, as
long as $f_{\rm{col}}$ and $\theta$ are relatively similar for both components, their
\textit{relative} radii should be fairly insensitive to the values adopted. On some level,
we might expect that $f_{\rm{col}}$ should be larger for the hotter component
(\citealt{Shimura95}), which would increase the radius of this component relative to the
cooler one. Large differences in $f_{\rm{col}}$ are not likely though (\citealt{Shimura95,
Davis06}), so we expect this will only influence the relative sizes inferred by up to a
factor of a few. In addition, if the \bb\ component does arise from the regions of the flow
interior to \rsp, then this will likely be subject to some level of geometric collimation by
the inner funnel. As such, the observed flux will be enhanced relative to the intrinsic flux
by a beaming factor $b$ (\ie $L_{\rm{int}} = bL_{\rm{obs}}$ adopting the nomenclature
of \citealt{King08}, where $b \leq 1$, and $b = 1$ represents isotropic emission). Since
the outer disc would be largely immune to similar geometric beaming this would serve
to decrease the radius of the hotter component relative to the cooler one, but only by a
factor of $b^{-1/2}$. For illustration, the beaming factor of $b \sim 0.1$ suggested by
\cite{King08} for neutron star accretors reaching $L_{\rm{X}} \sim 10^{40}$\,\ergps\
(similar to P13) would only change the relative radii of the thermal components by a
factor of a few. Furthermore, truncation of the disc by the magnetic field will reduce the
degree of geometric beaming relative to the case in which the magnetic field has a
negligible effect on the flow (as assumed by \citealt{King08}). Given that these two
issues effects act in opposition to each other, and may result in corrections of a similar
order, the conclusion that \rsp\ and \rmag\ are comparable would still appear to hold.

If \rmag\ is fairly similar to \rsp, then this could potentially explain why the hotter thermal
component in P13 does not require a strong range of temperatures. If the magnetic field
quickly truncates the disk for $R <$ \rsp, then the geometrically thick inner region of the
flow would only exist over a fairly narrow range of radii, and subsequently only emit over
a narrow range of temperatures. The preference for a single blackbody/steep radial
temperature index is likely driven in part by the assumption that the disc extends out to
radii $\gg$ \rin\ (taken to be \rmag\ in our simple picture), which is implicit in the
\diskpbb\ model. This could artificially force $p$ to be steep in the fits since this is the
only way to make the \diskpbb\ model effectively emit over a narrow range of
temperatures, even if in reality $p < 0.75$ between \rsp\ and \rmag, as would be
expected for a thick super-Eddington disc.

It is interesting to note that \cite{King17ulx} also suggest that \rsp\ and \rmag\ are
similar for the three ULX pulsars identified to date. However, we caution that this is
based on a model which assumes that the \textit{lower} temperature component
represents the blackbody emission from within the inner funnel of a super-Eddington
flow. Furthermore, the model is designed to reproduce a luminosity--temperature
relation of the form $L \propto T^{-4}$ (note the sign on the exponent), which has
previously been claimed observationally for this component. This is primarily based on
the work of \cite{Kajava09}, who find an anti-correlation between luminosity and
temperature with this approximate form by fitting the results for the soft  thermal
component compiled from a sample of ULXs. More recent analyses do not support
such a relation for the soft component when individual sources are considered
(\citealt{Miller13ulx}), instead finding positive indices (\ie luminosity positively correlated
with temperature). Critically, for P13 we find that both the cooler \diskbb\ and the hotter
blackbody components show positive correlations between luminosity and temperature,
and certainly do not follow $L \propto T^{-4}$. Nevertheless, our results would appear
broadly consistent with the broad conclusions of \cite{King17ulx} with regard to the
similarity of \rsp\ and \rmag.

Although it is challenging to constrain the absolute value of \rmag\ for P13 from our
analysis given the issues related to beaming and the colour correction discussed above,
if \rmag\ $<$ \rsp\ (as is supported by the requirement for two non-pulsing blackbody
components) then we can obtain a rough upper limit on the magnetic field of the neutron
star by taking the limit that \rmag\ = \rsp. For a dipole field and a typical neutron star
mass and radius, the magnetospheric radius is given by \rmag\ $\sim (2.6 \times 10^{8})
\dot{M}_{17}^{-2/7} B_{12}^{4/7}$ cm (\citealt{Lamb73, Cui97, Fuerst17}), where
$\dot{M}_{17}$ is the accretion rate at \rmag\ in units of $10^{17}$\,\gps\ and $B_{12}$ is
the magnetic field strength in units of $10^{12}$\,G.  Standard super-Eddington theory
suggests the location of \rsp\ is related to the input accretion rate:
$R_{\rm{sp}}/R_{\rm{G}} \sim (27/4)\dot{m}_{0}$ (independent of the nature/magnetic
configuration of the central accretor), where $\dot{m}_{0}$ is the input accretion rate in
Eddington units (\citealt{Shakura73}), so taking this limit allows us to estimate both \rmag\
and $\dot{M}_{17}$, assuming conservatively that the accretion rate at both \rsp\ and
\rmag\ is the same as $\dot{m}_{0}$ (in reality, we expect there to be strong mass loss
from the system; this should primarily occur interior to \rsp, being the point at which the
accretion flow becomes locally Eddington, and will reduce the accretion rate at \rmag\
relative to that at \rsp, which further emphasises that our calculation is an upper limit).

Our prior estimate of \rsp\ (Section \ref{sec_BB}) assumed both $\cos\theta$ and
$f_{\rm{col}} = 1$, to conservatively compare with neutron star radii. The former may be
reasonable, since we appear to have an unobscured view of the inner regions of the
flow (\citealt{Sutton13uls, Middleton15}), but the colour correction for a standard disc
(expected beyond \rsp) is more typically expected to be $f_{\rm{col}} \sim 1.7$
(\citealt{Shimura95}). Adopting this value for $f_{\rm{col}}$ increases the estimate for
\rsp\ to $\sim$900\,km, or equivalently $R_{\rm{sp}}/R_{\rm{G}} \sim 450$, and therefore
$\dot{m}_{0} \sim 66$. For a standard 1.4\,\msun\ neutron star the Eddington accretion
rate is $\dot{M}_{\rm{E}} = 2.2 \times 10^{18}$\,\gps\ (assuming a radiative efficiency of
$\eta \sim 0.09$), so the accretion rate at \rmag\ would be $\dot{M} \sim 1.4 \times
10^{20}$\,\gps. Substituting all this back into the original expression for \rmag\ and
rearranging, we therefore find that $B_{12} \lesssim 6$. Interestingly, the co-rotation
radius for P13 is also \rco\ $\sim$ 900\,km, so we have \rsp\ $\sim$ \rco. Since we know
that \rmag\ $<$ \rco, as accretion must be occurring in P13, this constraint would also
imply $B_{12} \lesssim 6$. Although we expect the similarity of \rsp\ and \rco\ is
coincidental (as long as \rmag\ $<$ \rco\ there is no reason \rco\ should be an important
radius in terms of the observed emission), this does provide further support for our
conclusion that \rmag\ $<$ \rsp, and that the thick inner disc does form over some range
of radii.

This is consistent with the estimate in \cite{Fuerst16p13}, who found $B_{12}
\sim 1.5$ based on the observed spin-up rate of the neutron star. However, we stress
that our calculation fundamentally depends on the association of the characteristic
radius of the cooler thermal component with \rsp. Furthermore, since $B \propto
R_{\rm{M}}^{7/4}$, any errors in the radius will be magnified in the magnetic field limit
estimated here. Our limit on $B$ therefore depends very strongly on the value assumed
for the colour correction factor, since the radii of the thermal components are themselves
proportional to $f_{\rm{col}}^{2}$. We also stress that \rmag\ is most sensitive to the
dipolar field, as any quadrupolar component will decay much more rapidly with radius, so
these constraints should be considered on the dipolar component only. A stronger
quadrupolar component to the magnetic field that only acts close to the neutron star, as
suggested by \cite{Israel17}, cannot be excluded by these considerations.

One of the basic predictions of super-Eddington accretion is that powerful winds
should be launched from the radiation-pressure dominated regions of the flow (\ie
primarily interior to \rsp; \eg\ \citealt{Poutanen07}). Indeed, unambiguous evidence for
powerful outflows has now been seen through resolved, blueshifted atomic features in a
handful of other ULX systems with deep \xmm\ coverage, primarily in the soft X-ray band
($<$2\,keV) but in the case of NGC\,1313 X-1 also in the iron \ka\ band
(\citealt{Pinto16nat, Pinto17, Walton16ufo}). If our proposed scenario for P13 is correct,
and thick inner regions of the flow do begin to form, then we might expect to see
evidence for a similar wind in P13, either through ionised absorption features if the
outflow crosses our line of sight, or ionised emission otherwise. Given the low levels of
neutral absorption here, this could be tested with future deep observations combining
either \xmm\ or \textit{XARM} (the \hitomi\ replacement mission; \citealt{ASTROH_tmp})
together with \nustar.

\subsection{The Low-Flux State}

In addition to the bright state observations, we also investigate the spectrum obtained
during the off-state seen between 2011 and 2013 (epochs C2, C3, C4 and X1). The
spectrum during these epochs is still very hard, and remains variable (Figure
\ref{fig_spec_ME}). However, simple powerlaw fits leave clear line-like residuals at
$\sim$1\,keV, and the fit is significantly improved with the addition of a plasma
component, which we argue is associated with the P13 system, rather than
galactic-scale plasma emission (e.g. star formation). If P13 does launch a strong wind
while at ULX luminosities, this plasma component could potentially arise from
collisionally-ionised material previously ejected from the accretion flow that becomes
visible with the extreme drop in flux from the central accretor.

The hard ($\Gamma \sim 0.7$) and variable powerlaw emission almost certainly
suggests that there is still some residual accretion onto the neutron star during these
off-states. After the discovery of pulsations from M82 X-2, \cite{Tsygankov16}
suggested that the low-flux states seen in this source could be related to the onset of
the propeller effect. This occurs when \rmag\ moves outside \rco, such that the
magnetic field ends up acting as a barrier to further accretion, rather than channelling
the material onto the neutron star poles. As such, for \rmag\ $>$ \rco\ the accretion rate,
and therefore the observed X-ray luminosity, suddenly drops precipitously. We do not
see any clear evidence for an accretion disk component in these data; if we replace the
\mekal\ component with \diskbb, the fit is better than the simple powerlaw model
(\chisq/DoF = 44/27), but not as good as the \mekal\ model, and the line-like residuals
at $\sim$1\,keV remain. This could suggest the disk would need to be truncated at a
sufficiently large radius that it makes a negligible contribution to the observed flux in the
X-ray band.

Nevertheless, the residual accretion flux is still quite substantial, with $L_{\rm{X}} \sim
3-6 \times 10^{37}$\,\ergps\ in the 0.3--10.0\,keV band, \ie $\sim$0.15--0.3
$L_{\rm{E}}$ for a 1.4\,\msun\ neutron star. This is quite similar to M\,82 X-2, which still
radiates at $\sim$10$^{38}$\,\ergps\ in its off-state (\citealt{Brightman16m82a}). As
noted by \cite{Tsygankov16}, this only requires a few percent of the accreting matter to
leak through the magnetosphere, which may not be unreasonable. However, there is
now evidence that the low-flux periods in M82 X-2 are related to its $\sim$62\,d
super-orbital cycle (Brightman et al. 2017, \textit{submitted}), which would call into
question the propeller interpretation for that system. If confirmed, this would establish
a significant difference between M82 X-2 and both NGC\,5907 ULX and P13, as the
timing and duration of the off-states seen in the latter two means they cannot be related
to their similar $\sim$64\,d and $\sim$78\,d periods. This means that there may not be
a single physical origin for these off-states common to all three systems, despite the
changes in observed flux being similar in all three cases, and so the propeller effect
cannot be excluded for the low-flux states in P13.

\subsection{Implications for the ULX Population}

The qualitative similarity of the broadband spectrum of P13 to the rest of the ULX
population has significant implications. Since P13 is a known super-Eddington accretor,
the \xmm+\nustar\ observation of this source presented here robustly demonstrates that
these spectra really are associated with super-Eddington accretion. This confirms the
conclusions of our previous work on broadband observations of ULXs
(\citealt{Bachetti13, Walton14hoIX, Walton15hoII, Walton17hoIX, Rana15, Mukherjee15}),
which were based on the lack of similarity between ULX spectra and the standard modes
of sub-Eddington accretion onto black holes (see \citealt{Remillard06rev} for a review),
rather than a well-established knowledge of the spectral appearance of super-Eddington
accretion at ULX luminosities.

The similarity between P13 and the broader ULX population may even suggest that
more of these sources could be powered by accreting neutron stars. Although
pulsations have not currently been observed from any other sources
(\citealt{Doroshenko15}), for M82 X-2 the pulsations are known to be transient
(\citealt{Bachetti14nat}), and this also appears to be the case for NGC\,5907 ULX
(\citealt{Israel17}). \cite{Pintore17} also suggest that additional ULXs could be pulsar
systems, arguing that ULX spectra are well fit by spectral models typically applied to
sub-Eddington pulsars: a low-energy blackbody (peaking below 2\,keV) and a
powerlaw with an exponential cutoff produced by an accretion column that dominates
at higher energies ($>$2\,keV). Our work suggests that the spectral decomposition
used by \cite{Pintore17} is not actually correct, and that the situation in P13 is
more complex as the 2--10\,keV emission does not exclusively arise from the column.
Instead, we find that this only reliably dominates the emission above 10\,keV, resulting
in the high-energy excess seen with models that only invoke two continuum
components (Figure \ref{fig_spec}). It is possible therefore that the presence of such a
hard excess is indicative that the accretion in the innermost regions occurs via
magnetically channelled columns, rather than via a disc. We stress again that a number
of other ULXs have now been seen to exhibit similar excesses above $\sim$10\,keV,
so if this is correct then neutron stars/pulsars could potentially be prevalent in the ULX
population.

Since the pulsations are seen to be transient in 2/3 of the known ULX pulsar systems,
additional means of firmly identifying pulsar/neutron star accretors may well be the key
to expanding this population. Indeed, \cite{King17ulx} suggest that pulsations may only
be observable from neutron star ULXs when \rsp\ and \rmag\ are similar (as we find
appears to be the case for P13), as they argue that for \rmag\ $\ll$ \rsp\ the pulsations
could be diluted to the point of being undetectable by the non-pulsed emission from the
disc. Although our work finds that the highest energies ($\gtrsim$15\,keV) should be
relatively immune to such issues as the column always dominates, the low count-rates
observed at these energies would still serve as a major barrier to the detection of
pulsations, further emphasising the need for additional means of identification.

One alternative method may be to search for other sources that show similar off-states
to those seen from all three ULX pulsars, as suggested by \cite{Fuerst16}. In addition to
this, based on our analyis of P13 in this work we tentatively suggest that finding that the
hotter thermal component does not require a run of temperatures may be another
indicator of a neutron star accretor. This would provide an indication that the disc is
truncated close to \rsp, despite the source maintaining ULX luminosities, and would in
turn imply a magnetized central accretor.

\section{Summary and Conclusions}
\label{sec_conc}

We have undertaken a detailed X-ray spectral analysis of the available data for P13 --
a key laboratory for ULX pulsars owing to its lack of source confusion, persistently
detected pulsations and low obscuration -- focusing on the first high-energy detection
of this source with \nustar. Through time-averaged, phase-resolved and multi-epoch
studies, we find that two thermal components are required to fit the data below 10\,keV,
in addition to a third continuum component above 10 keV to account for the
high-energy excess seen with dual-thermal models. This is qualitatively very similar to
the rest of the ULX population with broadband coverage, for which the nature of the
accretors generally remain unknown, confirming the super-Eddington nature of these
sources and suggesting that additional ULXs may be neutron stars/pulsars. Our
phase-resolved analysis finds that the third, high-energy component is associated with
the magnetically collimated accretion columns that must be present in P13 in order to
produce the observed pulsations. Neither of the two thermal components can arise
from the neutron star itself, as their characteristic radii at least an order of magnitude
too large, suggesting that they both arise from the accretion flow. In turn, this suggests
that the accretion flow has two distinct regions, as expected for super-Eddington
accretion flows which should transition from a standard thin outer disc to a
geometrically thick inner disc at the point that the flow reaches the local Eddington limit
(\rsp). However, the radii for these two components appear to be comparable, and the
hotter component is better described as a single blackbody than a thick,
advection-dominated accretion disc, which should have a strong run of temperatures.

We suggest that P13 may represent a scenario in which the magnetospheric radius --
the point at which the disk is truncated and the accreting material is forced along the
magnetic field lines -- is smaller than, but still similar to \rsp. This would allow the
thicker inner regions of the super-Eddington flow to begin to develop, before being
truncated prematurely by the magnetic field. Such a scenario would simultaneously
explain the need for two thermal components, their fairly similar radii, and the fact that
the hotter component does not need a run of temperatures, as the thick inner disk
would only persist over a fairly narrow range of radii, and thus only emit over a fairly
narrow range of temperatures. This similarity would imply an upper limit to the neutron
stars dipolar magnetic field of $B \lesssim 6 \times 10^{12}$\,G, albeit with large
uncertainties. If this is correct, evidence of similar truncation of the disc in other sources
could offer a means of identifying additional neutron star/pulsar ULXs, as this would not
be expected for black hole systems.

Finally, P13 is known to exhibit unusual `off' states in which the X-ray flux drops by
many orders of magnitude, potentially related to the propeller effect. We also examine
the spectra obtained during one of these events, and find that the data require both
a hard powerlaw component, suggesting residual accretion onto the neutron star, and
emission from a thermal plasma, which we argue is likely associated with the P13
system.

\section*{ACKNOWLEDGEMENTS}

The authors would like to thank the reviewer for the feedback provided, which helped
to improve the final version of the manuscript. DJW and MJM acknowledge support from
an STFC Ernest Rutherford fellowship, ACF acknowledges support from ERC Advanced
Grant 340442, and DB acknowledges financial support from the French Space Agency
(CNES). This research has made use of data obtained with \nustar, a project led by
Caltech, funded by NASA and managed by NASA/JPL, and has utilized the \nustardas\
software package, jointly developed by the ASDC (Italy) and Caltech (USA). This work
has also made use of data obtained with \xmm, an ESA science mission with
instruments and  contributions directly funded by ESA Member States. 


\bibliographystyle{/Users/dwalton/papers/mnras}

\bibliography{/Users/dwalton/papers/references}

\label{lastpage}

\end{document}